\documentclass[twocolumn]{aastex701}

\usepackage{siunitx}
\usepackage{physics}
\usepackage{graphicx}
\usepackage{subcaption}
\usepackage[whole]{bxcjkjatype}

\newcommand{\ph}{_{\rm ph}}

\usepackage{placeins}
\usepackage{lineno}

\graphicspath{{./}{figures/}}

\received{\today}

\shorttitle{TDE line profile}
\shortauthors{Takifuji \& Maeda}

\begin{document}
\linenumbers
\title{
Different origins for the line-emitting and photospheric components in Tidal Disruption Events: \\
A case for ASASSN-14ae}

\author[0009-0006-7153-5020]{Hare Takifuji}
\affiliation{Department of Astronomy, Kyoto University, 
Kitashirakawa-Oiwake-cho, Sakyo-ku, Kyoto, 606-8502, Japan}
\email[show]{takifuji@kusastro.kyoto-u.ac.jp}
\author[0000-0003-2611-7269]{Keiichi Maeda}
\affiliation{Department of Astronomy, Kyoto University, 
Kitashirakawa-Oiwake-cho, Sakyo-ku, Kyoto, 606-8502, Japan}
\email{keiichi.maeda@kusastro.kyoto-u.ac.jp}

\begin{abstract}
Tidal disruption events (TDEs) are believed to launch outflows that play a crucial role in producing optical emission lines. 
In this study, we investigate emission properties expected within the wind-driven outflow framework, in which an outflow is continuously injected from the central engine. We perform radiative transfer calculations to compute synthetic line properties and apply the model to the optical TDE ASASSN-14ae. 
We find that a two-component structure with a high density contrast, consisting of an optically thick inner layer and an optically thin outer layer, is required to simultaneously reproduce the observed photospheric and H$\alpha$ properties. The optically thin component exhibits a power-law temporal evolution in the mass-outflow rate and kinetic power, with the total ejected mass of $\sim0.25\,M_{\odot}$, indicating that it might originate in a super-Eddington accretion-disk–driven outflow. The optically thick component is responsible for the photospheric emission, and we argue that this component is more plausibly interpreted as a quasi-hydrostatic envelope rather than an outflow. Interestingly, the kinetic power of the optically-thin component and the bolometric luminosity originating from the optically-thick component are comparable, and they share the same temporal evolution. This finding indicates that these two components are energetically linked through an equipartition process -- pointing to a scenario in which the accretion energy is first converted to the kinetic energy of the outflow, and then this energy is dissipated to the envelope through the outflow-envelope interaction before it emerges from the surface of the envelope. 
\end{abstract}

\keywords{\uat{Tidal disruption}{112} --- \uat{Radiative transfer simulations}{2934} --- \uat{Supermassive black holes}{1444} --- \uat{Transient sources}{2779}}

\section{Introduction}

Tidal disruption events (TDEs) are energetic transients that occur when a star passes sufficiently close to a supermassive black hole (SMBH) at the center of a galaxy and is torn apart by tidal forces \citep{rees1988tidal,lacy1982naturea,loeb1997opticala}.  
The disrupted stellar material forms debris streams, a fraction of which falls back and accretes onto the SMBH \citep{evans1989tidala,loeb1997opticala,strubbeOpticalFlaresTidal2009,guillochon2015dark}.  
The rapid accretion phase can launch powerful outflows from the black hole–accretion disk system \citep{dai2018unified,metzger2016bright,metzger2022coolinga,jiangjiangyanfei2016prompt,hung2021discoverya}.  
As a consequence, TDEs are observed as luminous transient events at galactic nuclei \citep{lacy1982naturea,strubbeOpticalFlaresTidal2009}.

TDEs were first discovered in X-rays \citep{komossa1999discovery,komossa1999giant}.  
Only in the past $\sim$15 years have optically bright TDEs been routinely identified \citep{vanvelzen2011opticala}. 
With the rapid development of wide-field time-domain surveys, the number of optically selected TDEs has increased significantly \citep{vanvelzen2021seventeen,hammerstein2022final, nicholl2019tidal}.  
TDEs have been discovered and characterized by surveys such as intermediate Palomar Transient Factory  \citep[iPTF;][]{kulkarni2013intermediate}, Zwicky Transient Facility \citep[ZTF;][]{bellm2019zwicky}, and All-Sky Automated Survey for Supernovae \citep[ASAS-SN;][]{shappee2014man}.  
Furthermore, the Legacy Survey of Space and Time (LSST) conducted at the Vera C. Rubin Observatory is expected to substantially increase the TDE discovery rate in the coming years.

Optically selected TDEs exhibit diverse spectroscopic characteristics \citep{nicholl2020outflow,charalampopoulos2022detailed}.  
Their spectra commonly display broad emission lines of hydrogen and helium \citep{arcavi2014continuum,hammerstein2023integral}, and in some cases nitrogen-rich features \citep{leloudas2019spectral}. 
On the other hand, a subset of TDEs has been reported to show little or no prominent emission-line features \citep{hammerstein2022final,hammerstein2023integral}.
Despite this diversity, TDE line profiles share several common characteristics \citep{rothWhatSetsLine2018}; 
they typically exhibit broad emission components and weak or absent P-Cygni absorption features \citep{nicholl2020outflow}.

Several studies have attempted to reproduce TDE line profiles in order to probe the underlying outflow dynamics \citep{liu2017disc,cao2018large,rothWhatSetsLine2018,short2020tidal,parkinson2022optical}.  
Recently, \citet{zhangExploringPropertiesPhotosphere2024} investigated the properties of TDE photospheres and emission lines expected from a slim-disk wind model, and suggested that additional material outside the disk-wind photosphere, possibly supplied by debris self-collision during circularization, may be required to explain the observed line luminosities; the topic deserves further investigation based on a different approach, to see whether this is an issue for a specific model or a general problem and to identify the origin of the discrepancy.
In summary, the structure and origin of TDE-driven outflows remain poorly understood, and consequently the detailed formation mechanisms of the observed emission lines are still under debate. 

In this work, we investigate properties of the H${\alpha}$ emission expected in the wind-driven outflow framework; the outflow is assumed to be continuously injected from the `central engine'. The scenario successfully explains photospheric properties in TDEs \citep{uno2020winddrivena,piro2020windreprocessed, matsumoto2021limits}, 
while the line-formation process has been largely unexplored. We perform radiative transfer calculations to compute the resulting H$\alpha$ properties, and examine whether such a model can simultaneously reproduce the photospheric properties and H$\alpha$ properties observed for ASASSN-14ae as a representative example. 

This paper is organized as follows.
Sections~\ref{sec:MetOver} provides an overview of the methodology adopted in this study, followed by details of each physical process presented in Section~\ref{sec:methodofcalc}. 
Section~\ref{sec:gen_res} presents general emission properties expected in the wind-driven framework. The expected emission properties of the minimal, one-zone model are described in Section~\ref{sec:survey}, where we argue that such a model have critical difficulties to reproduce observed spectral properties of ASASSN-14ae as a prototypical example. We then introduce a two-component model in Section~\ref{sec:TC} and present the physical properties of these components derived for ASASSN-14ae. 
In Section~\ref{sec:discussion}, we discuss the origin(s) of these two components in a context of the TDE dynamics. The paper is closed in Section~\ref{sec:conclusions} with conclusions.

\section{Method Overview}\label{sec:MetOver}
In general, the wind-driven models \citep[e.g.,][]{uno2020winddrivena, piro2020windreprocessed}
take a phenomenological approach with which the outflow properties (e.g., the mass-outflow rate, $\dot M$ and the outflow velocity, $v$) are connected to the photospheric properties (the luminosity, $L_{\rm ph}$, the temperature, $T_{\rm ph}$, and the radius, $R_{\rm ph}$, as connected through the Stefan-Boltzmann relation). The structure above the photosphere is of interest in the present work, since this is the region where the lines form. 

Under the steady-state and spherically-symmetric approximations, our model is characterized by three parameters; $\dot{M}$, $v$, $L_{\rm ph}$. The model then returns the photospheric radius and temperature, as well as the radial structures of density and temperature above the photosphere. These are then used to compute the H$\alpha$ luminosity and profile through the radiation-transfer calculation. 
With the model output thus obtained, we explore what conditions are required to simultaneously reproduce the photospheric properties and H$\alpha$ properties of ASASSN-14ae as a representative case. 

We start with a `single-component' model, in which the photospheric properties and the line properties are computed based on the same model -- the model parameters are first constrained by the requirement to reproduce the photospheric properties, and then the predicted H$\alpha$ properties are compared with those of ASASSN-14ae at several epochs. We there show that this simplest approach indeed fails, clarifying a need to introduce a `two-component' model which is described as a combination of an optically-thick component (that forms the photosphere) and an optically-thin component (that provides a line-forming region above the photosphere). 

\section{Method of Calculations}\label{sec:methodofcalc}
\subsection{Outer wind structure}\label{sec:OWmodel}

For simplicity, we assume a constant-velocity field within the line-forming region at each epoch, as is frequently adopted within the steady-state description of a wind-driven model \citep[e.g.,][]{uno2020winddrivena}. This assumption does not require that the outflow velocity is fixed over the entire temporal evolution; in the application to ASASSN-14ae, the velocity is constrained independently at each epoch from the observed H$\alpha$ full width at half maximum (FWHM).
At each epoch, the velocity has a uniform magnitude throughout the outflow, and its direction is parallel to the radial vector. The density structure is thus described by the steady-state wind profile; 
\begin{equation}
        \rho = \frac{\dot M}{4\pi r^2 v} \ .
\end{equation}
We also assume a pure hydrogen composition for the outflow. 

The temperature structure above the photosphere is constructed assuming the free-streaming limit, rather than using the diffusion approximation frequently adopted in the previous works \citep[e.g.,][]{uno2020applicationa}. Omitting the contribution from the diffuse component, we assume that geometric dilution of the radiation from the photosphere dominates. We therefore adopt the following radial temperature profile: 
\begin{equation}
        T(r) = T_{\rm ph}\left(\frac{r}{R_{\rm ph}}\right)^{-1/2} \ .
\end{equation}

The photospheric radius is determined by the condition $\tau_{\rm eff}(R_{\rm ph}) = 1$, where the effective optical depth is given as 
\begin{equation}
    \tau_{\rm eff}(r) = \int_r^\infty \kappa_{\rm eff}(r')\,\rho(r')\,{\rm d}r' \ . 
\end{equation}
The effective opacity $\kappa_{\rm eff}$ is defined as 
\begin{equation}
    \kappa_{\rm eff} = \sqrt{3\kappa_{\rm a}(\kappa_{\rm s} + \kappa_{\rm a})} \ . 
\end{equation}
Here, $\kappa_{\rm a}$ is the absorptive opacity, given by
$\kappa_{\rm a} = 2.0\times10^{24}\,\rho\,T^{-7/2}$.
The electron-scattering opacity $\kappa_{\rm s}$ is expressed as
$\sigma_{\rm T} X_e / m_p$, where $\sigma_{\rm T}$ is the Thomson scattering cross section and $X_e$ is the ionization fraction.
Once the location of the photosphere is determined, we impose the Stefan–Boltzmann relation: 
\begin{equation}
    L_{\rm ph} = 4\pi R_{\rm ph}^2 \sigma_{\rm SB} T_{\rm ph}^4 \ .
\end{equation}

While we assume the steady-state description, we take into account the finite physical extent of the outflow set by the duration of the outflow since the launch. The outer boundary of the model is thus determined by $R_{\rm out} = \mathrm{day} \times v$. The parameter, $\mathrm{day}$, represents the observational epoch measured from the time when the outflow was launched. 

At each radius, the ionization fraction is calculated using the Saha equation \citep{georgeb.rybicki1985fundamentals}:
\begin{equation}
    \frac{n_e n_p}{n_{\rm H^0}}
    =
    \left(\frac{2\pi m_e kT}{h^2}\right)^{3/2}
    e^{-\chi_i/kT} \ ,
\end{equation}
where $n_e$, $n_p$, and $n_{\rm H^0}$ denote the number densities of electrons, protons, and neutral hydrogen, respectively, and $\chi_i \simeq 13.6\,\mathrm{eV}$ is the ionization energy of hydrogen.  
We assume ionization only from the ground state and adopt a statistical weight of 2 for neutral hydrogen. 
We discuss the validity and limitations of this LTE/Saha treatment in Appendix~\ref{sec:lte_validity}, where we shown that the recombination and photoionization timescales are much shorter than the expansion timescale; this justifies the use of the LTE/Saha treatment. Further we discuss there that introducing non-LTE treatment would not change the estimated values, e.g., mass outflow rate, more than a factor of two, and thus would not alter our main conclusions.

Similarly, the level populations of neutral hydrogen are computed by the Boltzmann distribution:
\begin{align}
    n_{n=2}
    =
    \frac{g_2}{g_1}\,n_{\rm H^0}
    \exp\!\left(-\frac{E_{21}}{k_{\rm B}T}\right)
    \simeq
    4\,n_{\rm H^0}
    \exp\!\left(-\frac{E_{21}}{k_{\rm B}T}\right) \ .
\end{align}
Here, the partition function is approximated by $g_1$, assuming that the ground level dominates the population.

\begin{figure}[t]
    \centering
    \includegraphics[width = \columnwidth]{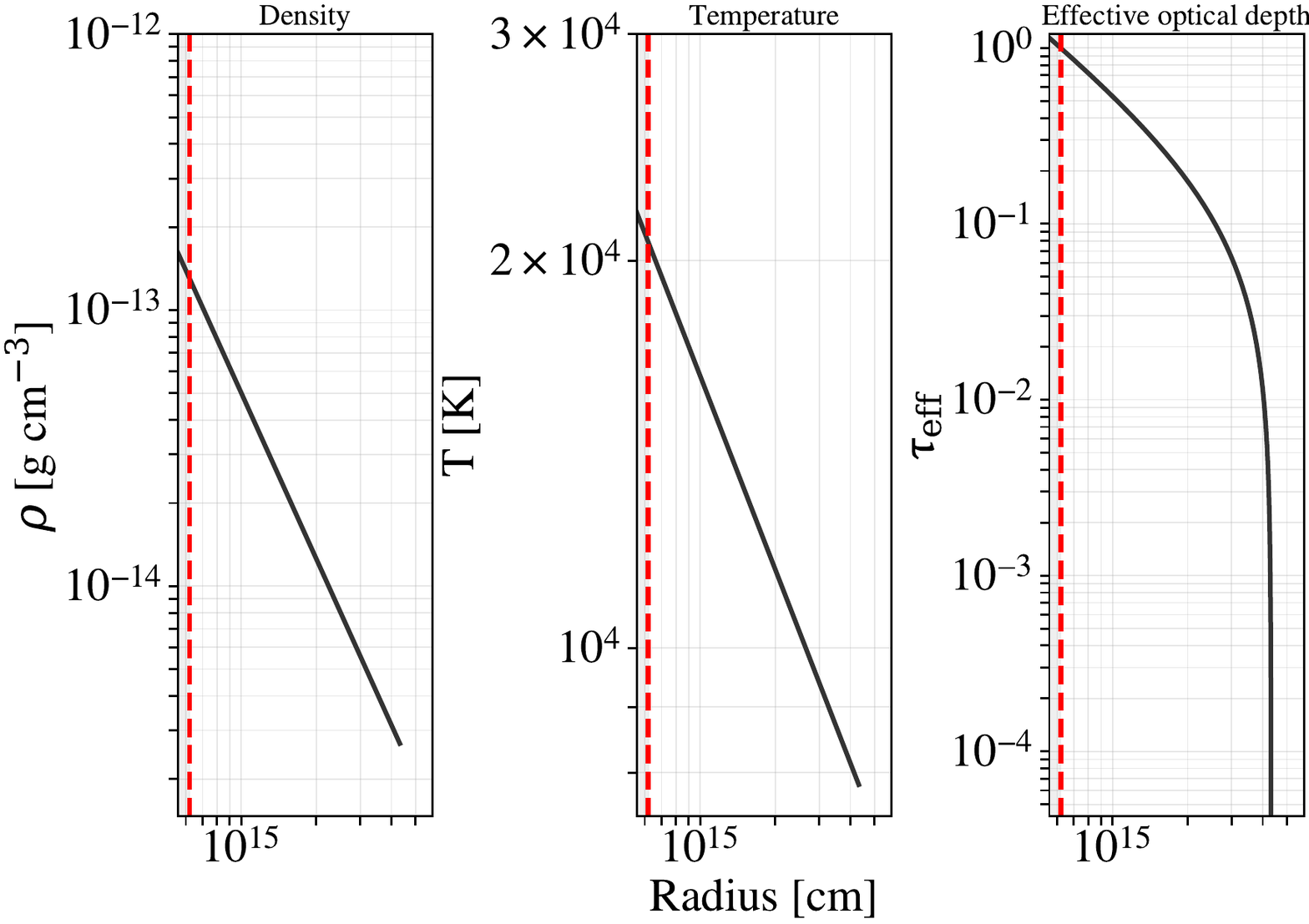}
    \caption{
    An example of the radial profiles of density (left), temperature (middle), and effective optical depth (right).
    The red dashed vertical line in each panel indicates the photospheric radius, defined by the condition $\tau_{\rm eff}=1$ in the right panel.
    }
    \label{fig:rho_temp_tau_wide}
\end{figure}

An example of the radial structure of the wind obtained through the above procedures is shown in Figure~\ref{fig:rho_temp_tau_wide}. In this particular example, the model parameters are set as follows: 
$\dot M=10.0 M_\odot \mathrm{yr}^{-1} ,v=1.0\times 10^9 \mathrm{cm}$ $\mathrm{s}^{-1}$, $L\ph=5.0\times 10^{43}\mathrm{erg}$ $\mathrm{s}^{-1}$, and $\mathrm{day}=50.0$ $\mathrm{days}$.

\subsection{Radiative Coefficients}\label{sec:rad_coe}

We consider two emission mechanisms for the H$\alpha$ line; resonant scattering and recombination emission. 
Under the Sobolev approximation \citep{sobolev1960moving,castor1970spectral,eastman1993spectruma} and 
the constant-velocity field, 
the Sobolev optical depth for the H$\alpha$ line is expressed as 
\begin{align}
\tau_{\rm sob}
= \frac{\pi e^2}{m_e c}\,
  f_{23}\,
  \lambda_0\,
  n_{n=2}\,
  \frac{r}{v\sin^2\theta}\,
  \left(1-e^{-h\nu_0/k_{\rm B}T}\right) \ .
\label{eq:sobolev}
\end{align}

Resonant scattering is evaluated under the thermalization approximation following \citet{kasen2006timedependent} \citep[see also][]{maedaSignaturesCompanionStar2014}:
\begin{equation}
    j_{\rm res}(r)=\alpha(r)\,S_\nu(r) \ .
\end{equation}
This approximation is equivalent to the limit of thermalization parameter $\epsilon = 1$ in \citet{castor1970spectral}.  
Here, $\alpha$ denotes the absorption coefficient and $S_\nu$ is the source function, which is approximated as $S_\nu = B_\nu(T)$ in this study. For the absorption coefficient, we adopt the expansion-opacity formalism of \citet{eastman1993spectruma}: 
\begin{align}
\alpha(r)
= \frac{\nu}{\Delta\nu}\,
  \frac{v}{cr}\,
  \sin^2\theta\,
  \left[1-\exp\!\left(-\tau_{\rm sob}\right)\right] \ .
\label{eq:alpha_exp}
\end{align}
We note that this expression is derived under the assumption of a constant-velocity field.

For recombination emission, we adopt the Case B approximation based on \citet{osterbrock1989astrophysics}:
\begin{equation}
    j_{\rm rec}(r)=\frac{h\nu_0 \,\alpha_{\rm eff}(T,n_e)\,n_e n_p}{4\pi} \ ,
\end{equation}
where $\alpha_{\rm eff}$ is the effective recombination coefficient.  
We use the tabulated values of $\alpha_{\rm eff}$ provided by \citet{osterbrock1989astrophysics}.  
Interpolation is performed with respect to temperature.  
Although the electron number densities in our model differ by orders of magnitude from the tabulated values, the dependence of $\alpha_{\rm eff}$ on $n_e$ is generally weak; therefore, we extrapolate the dependence on $n_e$. Photons emitted via recombination are locally trapped by absorption and scattering, and cannot escape immediately.  
This effect is evaluated using the escape probability $\beta$ \citep{castor1970spectral}:
\begin{equation}
    \beta = \frac{1-e^{-\tau_{\rm sob}}}{\tau_{\rm sob}} \ .
\label{eq:escape}
\end{equation}
The net recombination emissivity is then given by
\begin{equation}
    j_{\rm rec}^{\rm net}(r)=\beta(\tau_{\rm sob})\,j_{\rm rec}(r).
\end{equation}

The same formalism is applied to H$\beta$, using the corresponding atomic data and Case B effective recombination coefficient.

\subsection{Radiative transfer}\label{sec:rad_trans}
In this study, we implement a radiative transfer code based on the ray-tracing method, following \citet{suzuki20162d}; a large number of rays are propagated until they reach a projected virtual screen placed outside the outflow. At each point along a ray, the specific intensity is updated using the formal solution of the radiative transfer equation:
\begin{align}\label{eq:formal}
I_{\nu,\mathrm{out}}
= I_{\nu,\mathrm{in}}\,e^{-\Delta\tau_\nu}
+ S_\nu\left(1-e^{-\Delta\tau_\nu}\right) \ .
\end{align}
Here, $\Delta \tau_\nu = \alpha_\nu \Delta s$, where $\alpha_\nu$ is the absorption coefficient given by Equation~\ref{eq:alpha_exp}, and $\Delta s$ is the path length of each ray segment.
As a boundary condition, the rays intersecting the photosphere are injected at the photosphere with the input intensity given by $I_{\rm input}=B_\nu(T_{\rm ph})$, whereas the rays that do not intersect the photosphere are computed starting with the initial intensity of $I_{\rm input}=0$. 

Radiative transfer is treated under the Sobolev approximation \citep{sobolev1960moving,castor1970spectral}.
Within this framework, the photon emission and absorption occur only within the wavelength (or frequency) bin into which the rest-frame line wavelength is Doppler-shifted by the line-of-sight velocity of each cell. Accordingly, $I_\nu$, $\Delta \tau_\nu$, and $\alpha_\nu$ in Equation~\ref{eq:formal} are evaluated by considering only a single frequency bin determined by the local velocity at each spatial position. Similarly, the Planck function for the photons emitted from the photosphere is evaluated at the rest frame of the point considered. 

\section{General Properties}\label{sec:gen_res}
\begin{figure}[htbp]
    \centering
    
    \begin{subfigure}{\columnwidth}
        \centering
        \includegraphics[width=\linewidth]{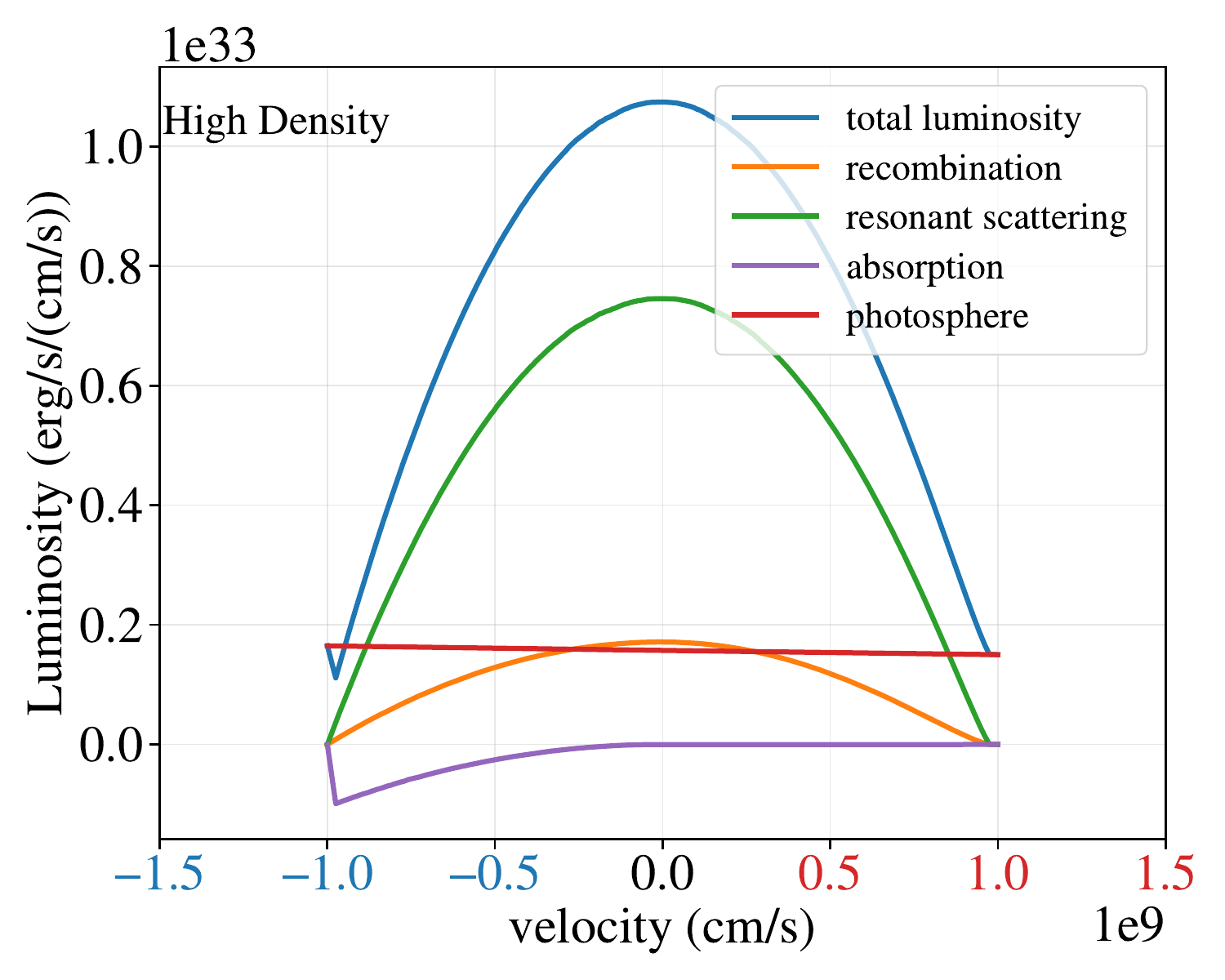  }
        \label{fig:LP_sample_high}
    \end{subfigure}
    
    \vspace{0.5em} 
    \begin{subfigure}{\columnwidth}
        \centering
        \includegraphics[width=\linewidth]{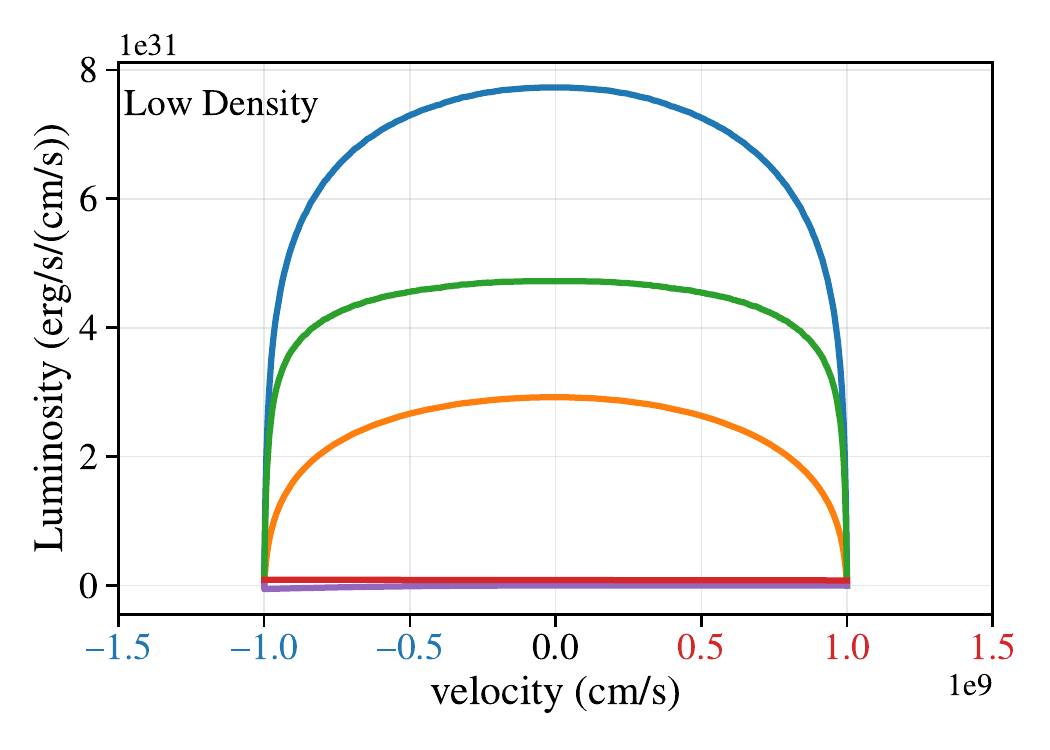}
        \label{fig:LP_sample_low}
    \end{subfigure}
    
    \caption{
    Calculated H$\alpha$ line profiles. Top: high-density model. Bottom: low-density.
    The orange, green, purple, red, and blue curves represent recombination emission, resonant scattering, absorption, input photospheric emission, and the total profile, respectively.}
    
    \label{fig:LP_sample}
\end{figure}

In this section, we present general properties of the optical emission for the wind-driven model, based on two specific parameter sets; 
$\dot M=10.0 M_\odot \mathrm{yr}^{-1}$ (high density model) and $\dot M=0.5 M_\odot \mathrm{yr}^{-1}$ (low density model). The remaining parameters are set as follows; $v=1.0\times 10^9 \mathrm{cm}$ $\mathrm{s}^{-1}$, $L\ph=5.0\times 10^{43}\mathrm{erg}$ $\mathrm{s}^{-1}$, $\mathrm{day}=50.0$ $\mathrm{days}$.
The radial distributions of the emissivities are shown in Figure~\ref{fig:jres_jrec}. 
Assuming spherically-symmetric outflow (see Appendix~\ref{sec:geom} for effects of asymmetry), the synthesized line profiles are shown in Figure~\ref{fig:LP_sample}, in which the decomposed contributions from individual emission components are shown together with the total line profile.

\begin{figure}[htbp]
    \centering
    \includegraphics[width=\columnwidth]{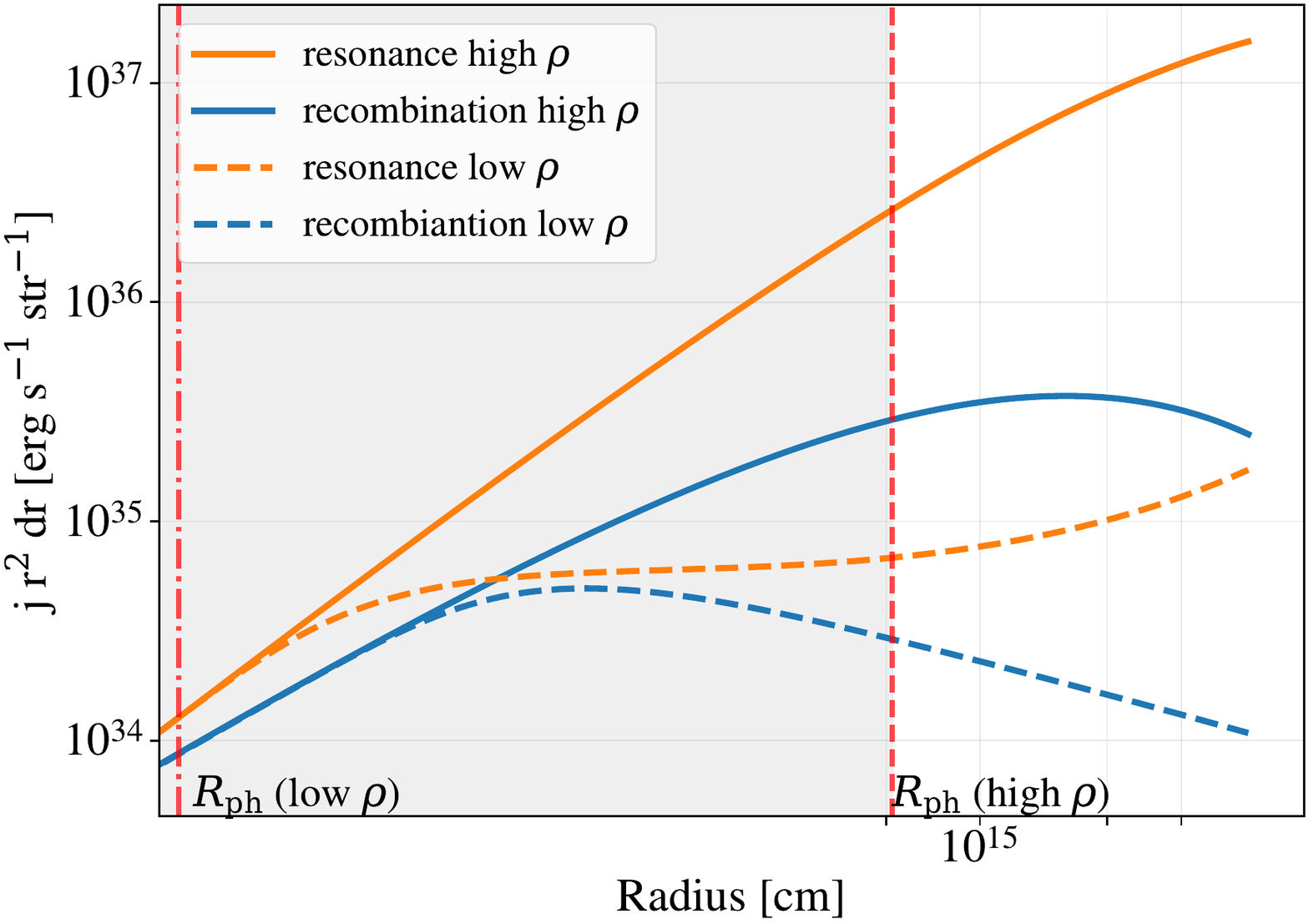}
    \caption{
    Radial profiles of the emissivities in two specific examples (see the main text for the model parameters). Orange curves represent resonant scattering, and blue curves represent recombination emission.
    Solid and dashed curves correspond to the high- and low-density models, respectively.
    Red-vertical dashed lines mark the photospheric radii.
    The emissivities are weighted by $r^2\,dr$ to represent the net radiative contribution from each spherical shell.
    }
    \label{fig:jres_jrec}
\end{figure}

Figure~\ref{fig:jres_jrec} shows that the overall emissivities are larger in the high–$\dot{M}$ case. In both models, resonant scattering dominates over recombination emission as the primary H$\alpha$ emission mechanism; 
this is because the large Sobolev optical depth leads to a low escape probability in Equation~\ref{eq:escape} and suppresses recombination emission, while enhancing the absorption of radiation and subsequent resonant emission.

Figure~\ref{fig:LP_sample} shows that absorption is weak compared with emission in both models. The resulting profile, weak P-Cygni and broad profile, reproduces the characteristic line profile commonly observed in TDEs \citep{rothWhatSetsLine2018}. The low-density model exhibits a flatter line profile than the high-density model; this behavior arises because the Sobolev optical depth above the photosphere falls below unity ($\tau_{\rm sob} < 1$) in the low-density model.

\section{Parameter Survey for ASASSN-14ae: One-component model}\label{sec:survey}
In this section, we perform radiative transfer calculations for steady-state, spherically symmetric outflows with a range of parameters. We aim to clarify whether the observed photospheric and line properties of ASASSN-14ae can be reproduced by the same model, i.e., by a `single-component' outflow. 

\subsection{Model Setting - Parameter Survey}\label{sec:survey_setting}
We adopt three key observational quantities as constraints; the photospheric radius ($R_{\rm ph}$), the photospheric temperature ($T_{\rm ph}$), and the H$\alpha$ line luminosity.  
The photospheric radius and temperature of ASASSN-14ae are taken from \citet{holoien2014asassn14ae}, while the H$\alpha$ line luminosities are adopted from \citet{charalampopoulos2022detailed}.  
Because the observational epochs reported in these two studies differ, we linearly interpolate the photospheric data of \citet{holoien2014asassn14ae} to match the epochs presented in \citet{charalampopoulos2022detailed}. 

Our model has three primary parameters; $\dot{M}$, $v$, and $L_{\rm ph}$. The combination of $R_{\rm ph}$ and $T_{\rm ph}$ (as observationally derived) gives the photospheric luminosity ($L_{\rm ph}$). 
The outflow velocity ($v$) is related to the full width at half maximum (FWHM) of the H$\alpha$ line reported in \citet{charalampopoulos2022detailed}.
Accordingly, only the mass-outflow rate $\dot{M}$ remains observationally unconstrained, and is treated as a free parameter; we vary it in the range of $0.1-40$ M$_\odot$ yr$^{-1}$. For each $\dot M$ values, we search for a proper value of $v$ which can reproduce the observed FWHM of H$\alpha$ emission line.
We target the following epochs for the model-observation comparison for ASASSN-14ae; 4, 30, 39, 51, 73, and 94 days since the discovery. 

We assume that ASASSN-14ae was discovered 50 days after the onset of the outflow expansion \citep{nicholl2020outflow,hung2017revisiting,holoien2020rise}. Accordingly, we define
\[
\mathrm{day} = \mathrm{epoch} + 50 \ .
\]
Here, $\mathrm{epoch}$ denotes the observational time since the discovery, while $\mathrm{day}$ represents the time used in our model calculations to define the geometric extension of the outflow (54, 80, 89, 101, 123, and 144 days for ASASSN-14ae).

\subsection{Results}\label{sec:result_epoch_survey}
\begin{figure}[t]
    \centering
    \includegraphics[width=\columnwidth]{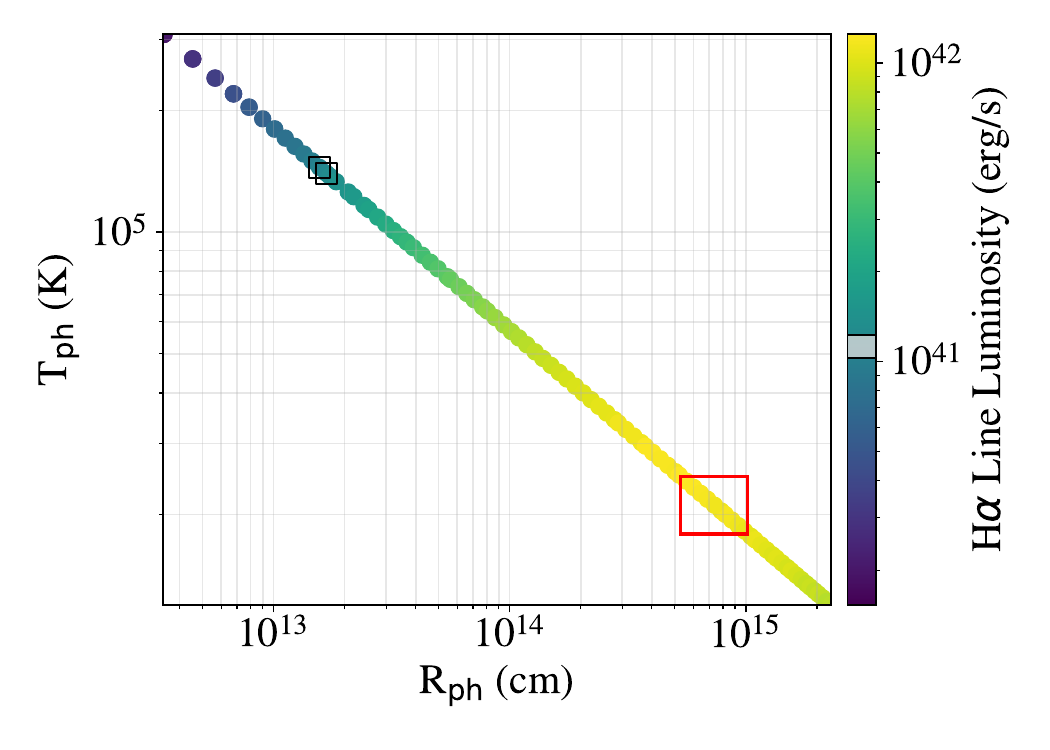}
    \caption{
    Results of the $\dot{M}$ survey at epoch 4d.
    Each point shows the predicted photospheric radius $R_{\rm ph}$ and temperature $T_{\rm ph}$ for given $\dot{M}$.
    The color scale represents the predicted H$\alpha$ line luminosity.
    The red square marks the region consistent with photospheric radius and temperature.
    Black squares indicate models consistent with the observed H$\alpha$ line luminosity.
    The white segment in the color bar highlights the luminosity range favored by the observations.
    }
    \label{fig:epoch4d}
\end{figure}

In our approach, we first derive the value of $\dot{M}$ that reproduces the observed photospheric radius and temperature at each epoch, as summarized in Table~\ref{tab:open}. Then, we investigate whether such a model consistently explains the observed H$\alpha$ luminosity; Figure~\ref{fig:epoch4d} presents the results for epoch 4d (54 days since the launch of the outflow); the corresponding plots for the other epochs are shown in Appendix~\ref{sec:other_epochs_single}.
 
As $\dot{M}$ increases, the photospheric radius increases. 
The mass-outflow rates favored by the photospheric constraints predict line luminosities that are significantly larger than observed (Figure~\ref{fig:epoch4d}) -- the observed photospheric properties favor high $\dot M$, while the line luminosity is consistent only with low $\dot M$. 
Consequently, no single value of $\dot{M}$ can simultaneously reproduce both the photospheric properties and the observed line luminosity.
We reach the same conclusion for all other epochs (see Appendix~\ref{sec:other_epochs_single}): there exists no viable $\dot{M}$ that satisfies both constraints within the single-component framework.

One might argue that introducing a geometric reduction of the emitting volume (e.g., via aspherical geometry through a collimated outflow) could suppress the line luminosity while preserving the photospheric properties. However, such geometric adjustments fail to resolve the tension. Reducing the effective emitting volume sufficiently to match the observed line luminosity leads to line profiles whose full width at half maximum deviates from the observed values (see Appendix~\ref{sec:geom} for details). Therefore, geometry alone cannot reconcile the discrepancy within a single-component outflow model.

\begin{deluxetable}{lc}
\tablecaption{$\dot M$ values that reproduce the observed photospheric properties, within the one-component model framework.\label{tab:open} 
}
\tablehead{
\colhead{Epoch (days)} &
\colhead{$\dot{M}$ ($M_\odot$ yr$^{-1}$)} 
}
\startdata
4 & 9.94 \\
30 & 8.81 \\
39 & 6.13 \\
51 & 4.01 \\
73 & 2.06 \\
94 & 1.06 \\
\enddata
\end{deluxetable}

\section{The Two-Component Outflow in ASASSN-14ae}\label{sec:TC}

\subsection{Model Setting}\label{sec:model_TC}
\begin{figure}[t]
    \centering
    \includegraphics[width=\columnwidth]{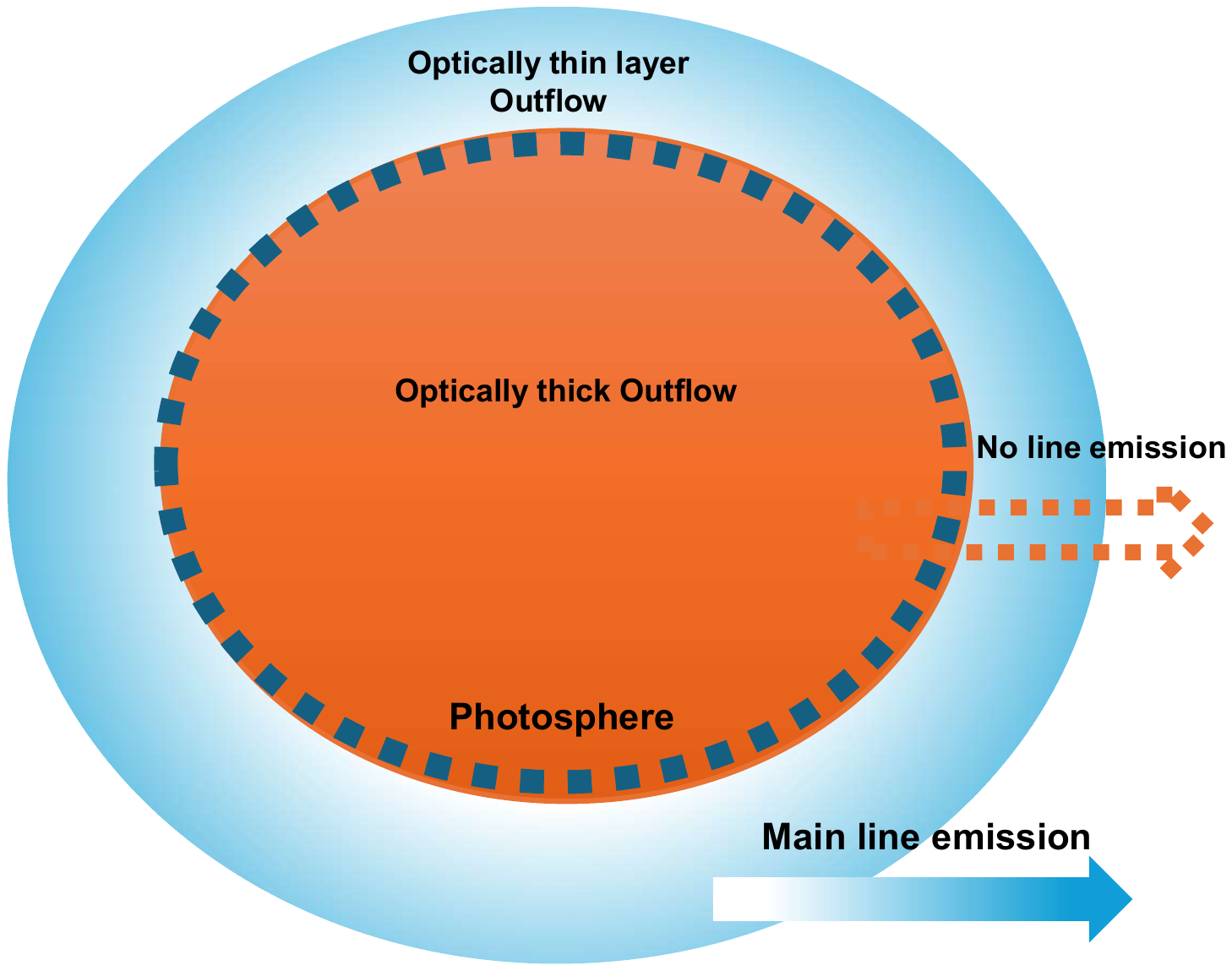}
    \caption{
    Schematic illustration of the two-component outflow model.
    The inner red region represents the high-density, optically thick component that sets the photospheric radius.
    The outer blue region represents the low-density, optically thin component responsible for line emission and luminosity.
    The two components share a common photosphere, indicated by the blue dashed line.
    Because the inner component is optically thick, the photosphere coincides with its outer boundary and contributes negligibly to line emission.
    }
    \label{fig:two_component}
    
\end{figure}

Given that there is no value of $\dot{M}$ which can simultaneously reproduce the observed photospheric radius, temperature, and line luminosity of ASASSN-14ae within a spherically symmetric single-component outflow framework, 
we introduce a two-component outflow model. The concept is illustrated in Figure~\ref{fig:two_component}.

In the two-component model, the outflow is divided into two layers; an inner high-density component and an outer low-density component. We note that we treat the configuration under spherical symmetry, while we do not reject a moderately collimated outflow (see Appendix~\ref{sec:geom}). The inner, optically thick outflow is introduced so that the photosphere is formed near the outer edge of this component. Namely, this component has very little material above the photosphere to contribute to the line emission.

The outer optically thin outflow is responsible for producing the line emission. This component does not contribute to the formation of the photosphere as a result of the low density. We assume that the photosphere is shared by the two components; this is possible since the boundary between the two components (i.e., the photosphere) has a huge jump in the density leading to the rapid transition from the optically thick to thin condition. 

For the optically thick component, the photospheric radius and luminosity ($R_{\rm ph}$, $L_{\rm ph}$) are directly specified by observations, as described in Section~\ref{sec:survey_setting}.  
The wind velocity $v$ is then converted from the outer radius of the outflow, $R_{\rm out}$ ($\sim R_{\rm ph}$), by subtracting this by the time since the outflow launch\footnote{We note that the velocity of the optically-thick component is thus not directly constrained by the observation, and this component may indeed represent a hydrostatic envelope rather than an outflow (see Section~\ref{sec:discussion} for discussion on this point).}. 
Consequently, the mass-outflow rate $\dot{M}$ alone is the free parameter for the optically thick component.
We vary $\dot{M}$ and examine how the photospheric radius approaches the outer radius ($R_{\rm out}=v\times \mathrm{day}$), thereby determining the lower limit of $\dot{M}$ required to reproduce the observed photosphere.

Since $R_{\rm ph}$ and $L_{\rm ph}$ are already fixed by the optically thick component, they do not constrain the properties of the optically thin component. 
For the optically thin component, the wind velocity $v$ is constrained by the observed FWHM as the same way in Section. \ref{sec:survey_setting}.
Thus, $\dot{M}$ is again the only free parameter in the optically thin component.
Since the photosphere is already determined by the optically thick outflow, the outer structure of the optically thin component (i.e., the line-forming region) is computed without solving for the photospheric location self-consistently.
We then vary $\dot{M}$ and identify the value that reproduces the observed line luminosity at each epoch.

\subsection{Results}\label{sec:result_TC}

\subsubsection{Optically thick component}\label{sec:result_thick}
\begin{figure}[t]
        \centering
        \includegraphics[width=\columnwidth]{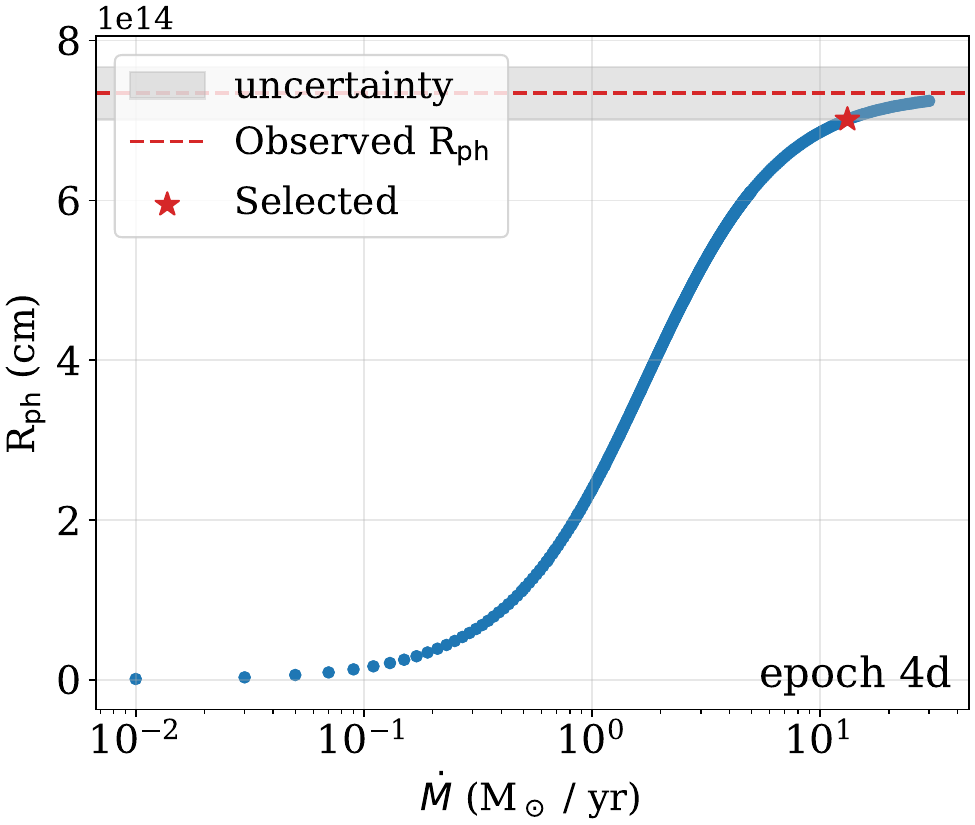}
        \caption{
        Dependence of the photospheric radius on the mass-outflow rate $\dot{M}$ for the optically thick component at epoch 4d.
        As $\dot{M}$ increases, the optical depth of the outflow rises, causing the photospheric radius $R_{\rm ph}$ to expand outward.
        The horizontal red dashed line marks the observed photospheric radius, and the gray shaded region represents its observational uncertainty.
        }
        \label{fig:ph_epoch4}
\end{figure}

The results of the $\dot{M}$ survey for the optically thick component at epoch 4d are shown in Figure~\ref{fig:ph_epoch4}.
As $\dot{M}$ increases, the density of the outflow increases, causing the photospheric radius to move outward.
Because the outer radius is fixed to the observed photospheric radius ($R_{\rm out}=R_{\rm ph,\,observed}$), the model-predicted photospheric radius $R_{\rm ph}$ asymptotically approaches the observed value as $\dot{M}$ increases.

We adopt the $\dot{M}$ values at which $R_{\rm ph}$ enters the observational error range as the lower limits for the optically thick component,  
under the assumption that this component would represent an `outflow'. This value should not be regarded as a direct measurement of the mass-ejection history, especially if this component instead represents a quasi-hydrostatic envelope as discussed in Section~\ref{sec:disc_thick}.
For epoch 4d, the predicted photospheric radius falls within the observational uncertainty for $\dot{M} \gtrsim 13.2\,M_\odot\,{\rm yr^{-1}}$ (Figure~\ref{fig:ph_epoch4}).
The lower limits for all the epochs are summarized in Table~\ref{tab:ph_LL_tune}.
In addition, we estimate the kinetic energy injection rate of the outflow using
\[
\dot{E} = \frac{1}{2}\dot{M}v^2. 
\]
Assuming that the optically thick component represents an outflow, the lower limits of the kinetic energy content as derived at different epochs are also listed in Table~\ref{tab:ph_LL_tune}. 

\subsubsection{Optically thin component}\label{sec:result_thin}
\begin{figure}[t]
        \centering
        \includegraphics[width=\columnwidth]{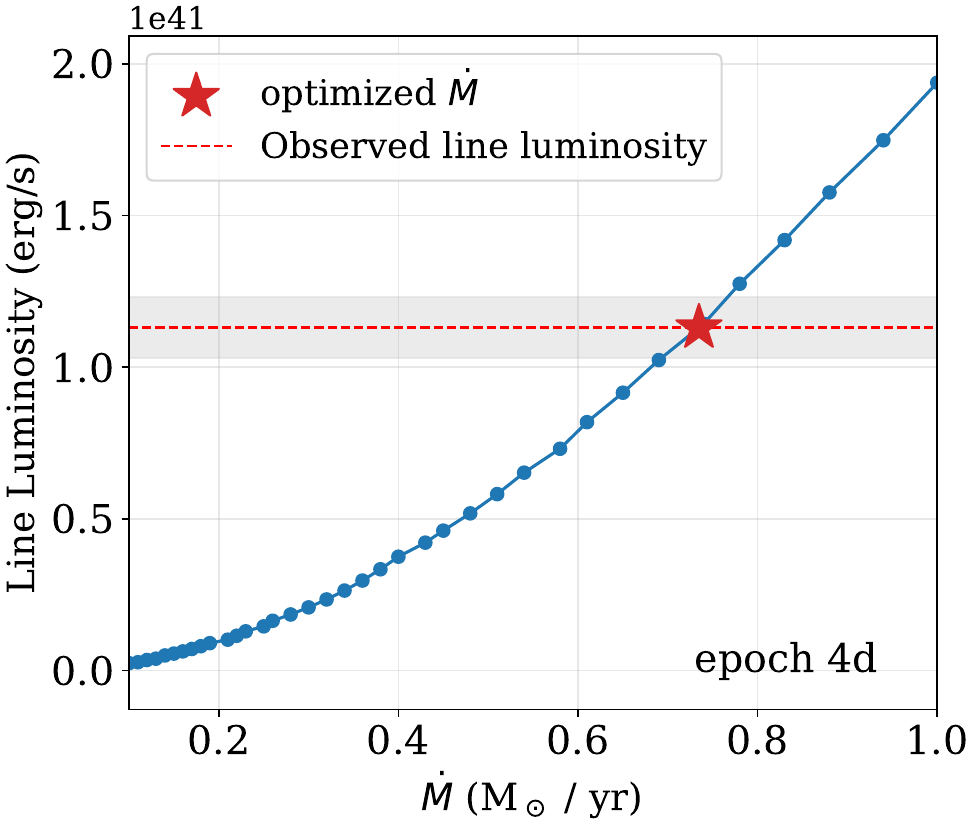}
        \caption{
        Dependence of the line luminosity on the mass-outflow rate $\dot M$ for the optically thin component at epoch 4d.
        As $\dot M$ increases, the line luminosity rises.
        The horizontal red dashed line marks the observed line luminosity, and the gray shaded region represents its observational uncertainty.
        }
        \label{fig:LL_epoch4}
        
\end{figure}

The results of the $\dot{M}$ survey for the optically thin component at epoch 4d are shown in Figure~\ref{fig:LL_epoch4}.
As $\dot{M}$ increases, the H$\alpha$ luminosity increases and reaches the observed value at a specific mass-outflow rate.
For epoch 4d, a value of $\dot{M} = 0.73\,M_\odot\,{\rm yr^{-1}}$ reproduces the observed line luminosity.

Table~\ref{tab:ph_LL_tune} summarizes the mass-outflow rates derived for the optically thin component across all epochs of ASASSN-14ae, including epoch 4d. We also list the kinetic energy injection rates $\dot{E}$, evaluated in the same manner as described in Section~\ref{sec:result_thick}.

\begin{deluxetable*}{lccccc}
    \tablecaption{$\dot M$ and $\dot E$ values in optically thick and thin outflows at each epoch in ASASSN-14ae\label{tab:ph_LL_tune}}
    \tablehead{
    \multicolumn{1}{c}{epoch} &
    \multicolumn{2}{c}{Optically thick} &
    \multicolumn{2}{c}{Optically thin} \\
    \cline{2-3} \cline {4-5}
    &
    \colhead{$\dot{M}$ ($M_\odot$ yr$^{-1}$)} &
    \colhead{$\dot E$ ($10^{42}$erg s$^{-1}$)}&
    \colhead{$\dot{M}$ ($M_\odot$ yr$^{-1}$)} &
    \colhead{$\dot E$ ($10^{42}$erg s$^{-1}$)}
    }
    \startdata
    4 & $\geq$ 13.2 & $\geq$ 10.3 & 0.73 & 10.4\\
    30 & $\geq$ 4.6 & $\geq$ 1.96  & 0.69 & 12.5 \\
    39 & $\geq$ 2.4 & $\geq$ 0.54 & 0.53  & 7.9\\
    51 & $\geq$ 0.95 & $\geq$ 9.3 $\times 10^{-2}$ & 0.42 & 5.5 \\
    73 & $\geq$ 0.24 & $\geq$ 6.3 $\times 10^{-3}$ & 0.29  & 3.2\\
    94 & $\geq$ 0.070 & $\geq$ 4.9 $\times 10^{-4}$ & 0.18 & 1.3 \\
    \enddata

    The tabulated values are obtained from independent steady-state calculations at each epoch. Their time-dependent interpretation is discussed in Section~\ref{sec}.
\end{deluxetable*}

\subsubsection{H$\beta$ luminosities}\label{sec}

To test the model with an additional line, we calculate the H$\beta$ luminosity using the same optically thin component parameters listed in Table~\ref{tab:ph_LL_tune}. These parameters were determined from the photospheric properties as combined with the H$\alpha$ luminosity and FWHM; H$\beta$ is not used to adjust any model parameter. Since the velocity is constrained by the H$\alpha$ FWHM, we focus here on the H$\beta$ luminosity, rather than separately fitting the H$\beta$ line width.

\begin{figure}[t]\centering\includegraphics[width=\columnwidth]{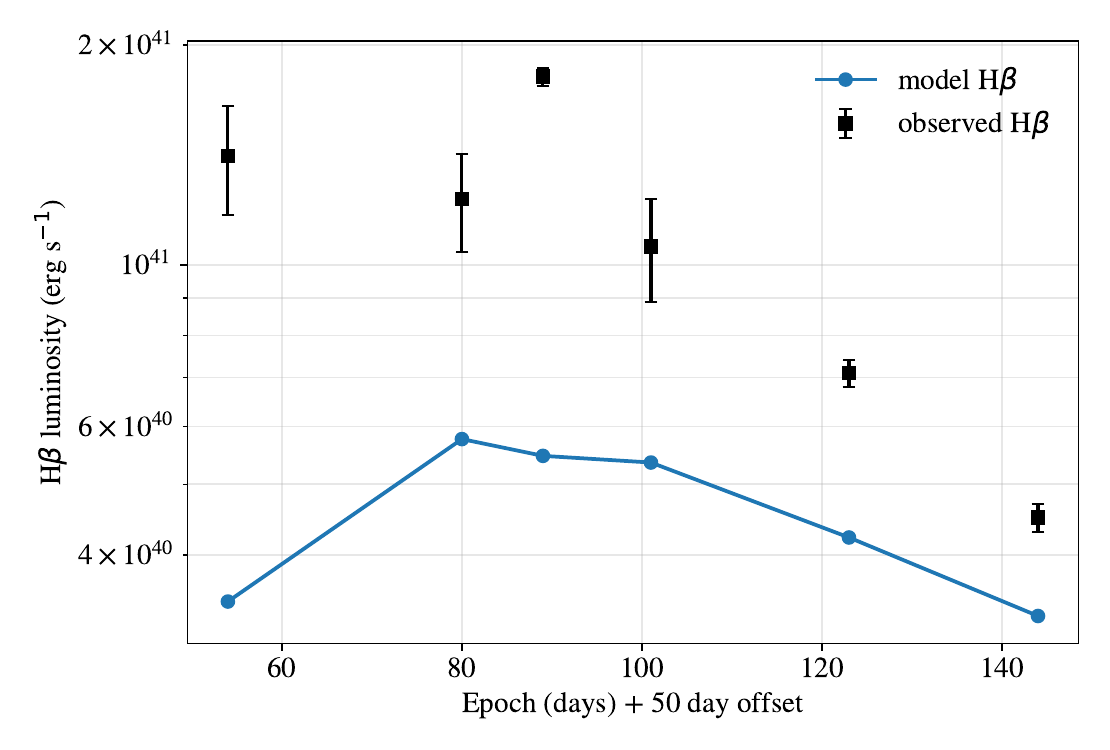}\caption{
Time evolution of the H$\beta$ luminosity (top) and the Balmer decrement, $L_{\rm H\alpha}/L_{\rm H\beta}$ (bottom), for ASASSN-14ae. The H$\beta$ predictions use the same optically thin component parameters fixed by the photospheric properties and H$\alpha$ at each epoch; no parameter is adjusted using H$\beta$.}\label{fig}
\end{figure}

Figure~\ref{fig} compares the predicted H$\beta$ luminosity with the observed ones at various epochs. The model generally underpredicts the H$\beta$ luminosity with the difference by a factor of a few in the first epoch and a factor of two thereafter. Given the simplified assumptions in the radiation transfer (e.g., the LTE assumption) which might result in quantitative uncertainty by a factor of two or a few, we regard that this level of agreement supports the general results and conclusions of the present work. 
Indeed, the discrepancy may suggest possible improvement in the radiation transfer treatment in the future. A key can be the low value of the Balmer decrement, found in the observations ($\sim 2$) as compared to the predicted ratio ($\sim 3-4$). The resonance line dominates in our calculations, but then the model would not produce the Balmer ratio smaller than $\sim 3$, under the full-thermalization assumption adopted in the present work. 
Accounting for non-LTE effects is expected to affect the H$\alpha$ line luminosity and, consequently, the inferred mass-loss rate $\dot{M}$; the resulting change in $\dot{M}$ is estimated to remain within a factor of approximately 3 (see Appendix D).
Further exploration needs non-LTE treatment of atomic populations and precise treatment of non-thermalized scattering in radiative transfer, which we postpone to the future.

\subsubsection{Line Profile}\label{sec:LP_TC}

While the modeled profile is somewhat flatter than the observed one, the model simultaneously reproduces the observed photospheric radius, temperature, line luminosity, and the FWHM of H$\alpha$. 
Results for the remaining epochs are shown in Appendix~\ref{sec:other_TC} (see also Figure~\ref{fig:lp_double_outflow_2x3_singlecol}), which illustrates the time evolution of the line profiles under the same procedure.

\begin{figure}[t]
        \centering
        \includegraphics[width=\columnwidth]{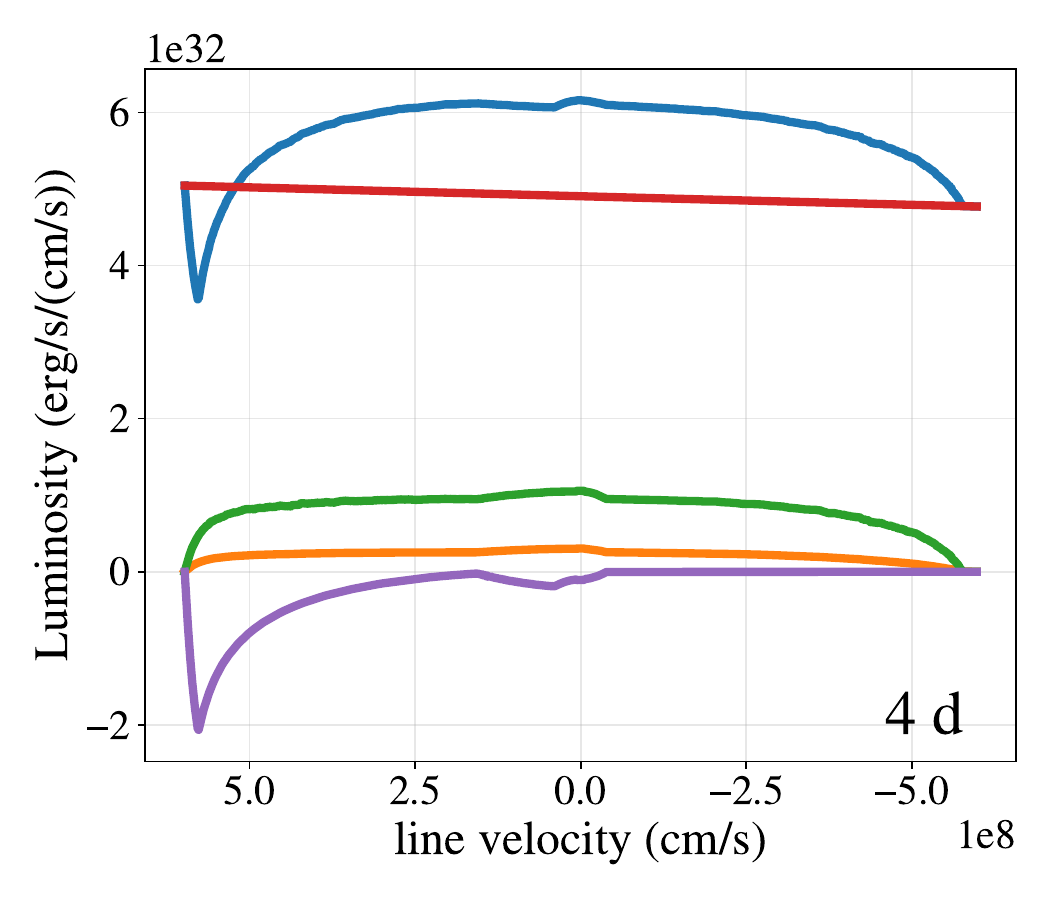}
        \caption{
        The synthetic line profile for the two component model at epoch 4d.
        The profile is decomposed to different contributions (see the caption of Figure~\ref{fig:LP_sample}).
        }
        \label{fig:LP_epoch4}
\end{figure}

Figure~\ref{fig:LP_epoch4} presents the emergent H$\alpha$ line profile computed by the radiative transfer calculations under the spherically symmetric, two-component outflow configuration at epoch 4d (see Table~\ref{tab:ph_LL_tune} for the model parameters). 
In our modeling, the outflow velocity mainly sets the velocity scale of the line profile, and is constrained at each epoch by the observed FWHM of H$\alpha$. Therefore, the agreement in FWHM should be regarded as a constraint imposed on the model rather than an independent prediction. The detailed profile shape, on the other hand, is not determined solely by the velocity; it also depends on the optical-depth structure, the relative contributions of resonant scattering and recombination emission, and the geometry of the line-forming region.
While the modeled profile is somewhat flatter than the observed one, the model simultaneously reproduces the observed photospheric radius, temperature, line luminosity, and the FWHM of H$\alpha$. 
Results for the remaining epochs are shown in Appendix~\ref{sec:other_TC} (see also Figure~\ref{fig:lp_double_outflow_2x3_singlecol}), which illustrates the time evolution of the line profiles under the same procedure.

\subsubsection{Time evolution of the mass-outflow rate and energetics}\label{sec:massloss}

\begin{figure}[t]
    \centering
    \begin{subfigure}{\columnwidth}
        \centering
        \includegraphics[width=\linewidth]{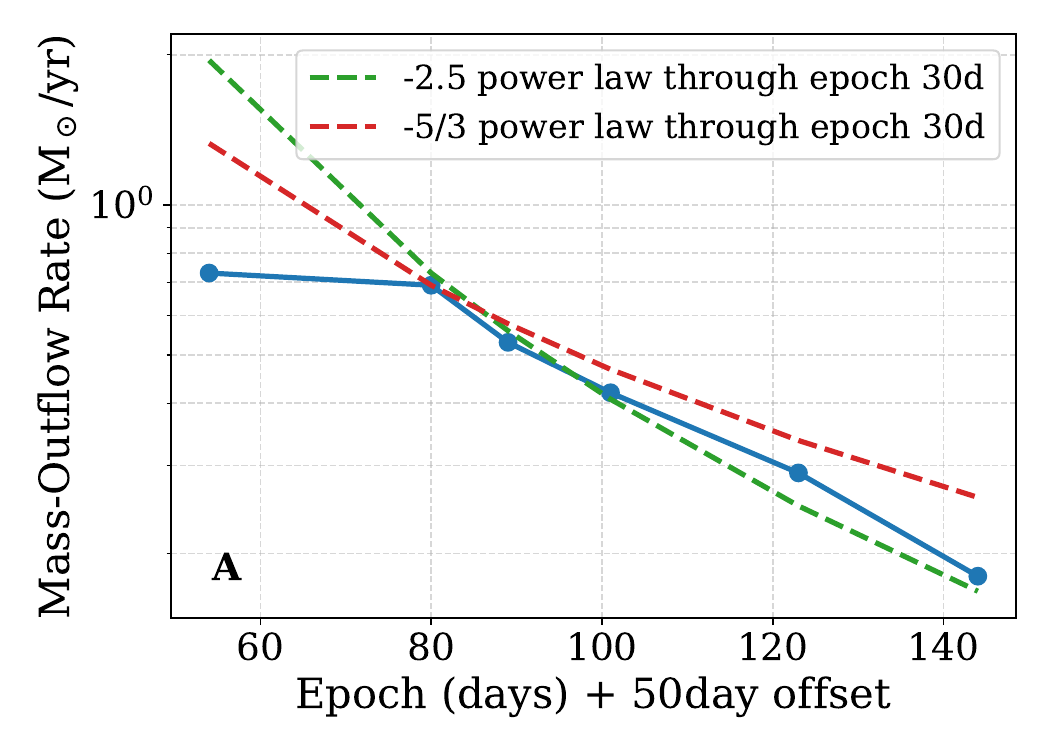}
        \label{fig:evol_Mdot}
    \end{subfigure}
    
    \vspace{0.5em} 
    \begin{subfigure}{\columnwidth}
        \centering
        \includegraphics[width=\linewidth]{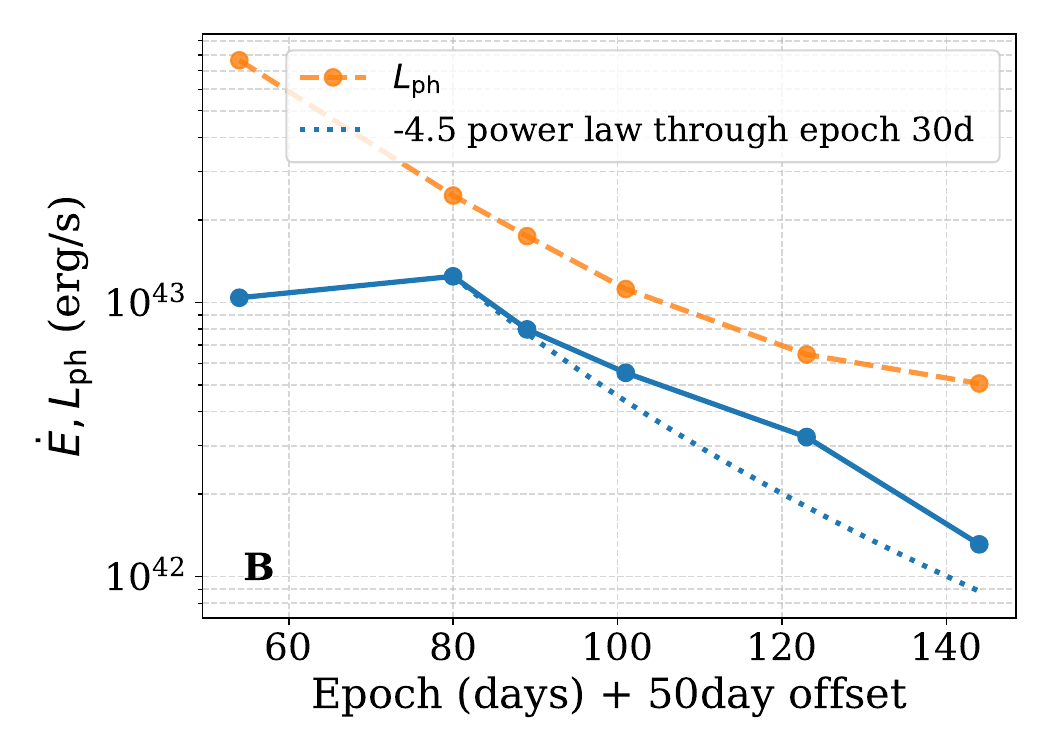}
        \label{fig:evol_kin}
    \end{subfigure}
    
    \caption{
        (A) The derived $\dot M$ values in the optically thin component for ASASSN-14ae.
        The time here is measured from the onset of the outflow.
        The red dashed curve denotes the $t^{-5/3}$ power law behavior, while the green one denotes $\propto t^{-2.5}$.
        (B) The same with A but for $\dot E$.
        The blue dotted curve denotes the $t^{-4.5}$ power-law behavior.
        The orange dashed curve denotes the evolution of $L\ph$ (i.e., the bolemetric light curve) of ASASSN-14ae.}
    
    \label{fig:time_evol}
\end{figure}

Figure~\ref{fig:time_evol} shows the time evolution of the mass-outflow rate ($\dot{M}$) and kinetic energy injection rate ($\dot E$) for the optically thin component (see Table~\ref{tab:ph_LL_tune}).
The evolution is well described by a power-law behavior following $\dot M \propto t^{-2.5}$, steeper than the canonical fallback-rate scaling $\dot M \propto t^{-5/3}$ often postulated for TDEs. 
The difference in the mass-outflow rate and the fallback rate is not surprising; the fallback rate does not necessarily trace the accretion rate, and further the fraction of the outflow to the accretion or fallback material may likely be time dependent -- this is indeed we will discuss in Section~\ref{sec:discussion}. 

The evolution of $\dot E$ is also well described by a power-law behavior but with a different index; this is largely consistent with $\dot E \propto t^{-4.5}$ (see Section~\ref{sec:discussion}). 
Interestingly, the kinetic power and its evolution roughly overlap with those seen in the radiated luminosity. 
We will discuss possible implications and caveats of these behaviors in Section~\ref{sec:discussion}. 

\section{Discussion}\label{sec:discussion}
We find that the observed photospheric properties and line properties can be reproduced (only) by separating the emitting region into optically thick and optically thin components (Section \ref{sec:result_TC}). 
Furthermore, the $\dot{M}$ values required to reproduce the observations differ significantly between the two components (see also Table~\ref{tab:ph_LL_tune}).  
In this section, we discuss the possible physical origins of these two components and their inferred properties.
Conversely, this provides a sanity check of our model, by investigating whether derived properties can be interpreted by physically-motivated scenarios. 

\subsection{Time-dependent interpretation of the two components}\label{sec}

The quantities listed in Table~\ref{tab:ph_LL_tune} are derived from steady-state calculations separately performed for different epochs.  The results thus directly trace the effective density normalization required near the photosphere and in the line-forming region at the corresponding epoch, but 
do not directly measure Lagrangian mass-loss histories. 
Therefore, it should be taken in mind that connecting the series of the steady-state solutions to infer the time-dependent history as we investigate in this section involves `interpretation'. 
This is especially important for the optically thick component; the lower limits for the outflow rate of this component (Table~\ref{tab:ph_LL_tune}) are derived assuming it would be an outflow, whereas Section~\ref{sec:disc_thick} argues that it is more plausibly a quasi-hydrostatic envelope. Further, there is no direct constraint on the velocity of the optically thick component, which makes the discussion on the mass-outflow rate itself meaningless in case it is not an outflow. Consequently, the factor of $\sim 20$ contrast between the two components at epoch 4d, and the fact that the optically thin value exceeds the formal lower limit for the optically thick component at epoch 94d, should not be interpreted as any transition in physical processes to produce the outflow(s). Within the present model, the robust conclusion is that a large density contrast between the photospheric and line-forming regions is required at every epoch.

A time-dependent physical configuration could nevertheless connect the two regions. Material launched into an outflow more strongly at early times would propagate outward, and its density and optical depth would decrease as it expands; such material could contribute to the optically thin line-forming atmosphere at later epochs. At the same time, the inner optically thick envelope can cool and contract, causing the photosphere to recede. Our present steady-state treatment does not retain the radial distribution of shells ejected at earlier epochs, and thus cannot determine this mapping quantitatively. A time-dependent radiation-hydrodynamic calculation of the coupled envelope and outflow will be required to follow this evolution self-consistently, which we postpone to the future.

\subsection{Origin of the optically thin outflow}\label{sec:TC_discuss}

Over the 140 days following the onset of the event, optically thin component of ASASSN-14ae ejects a total mass of $\sim 0.2\,M_\odot$ and a total kinetic energy of $\sim 1.0\times10^{50}\,\mathrm{erg}$. These values are obtained by integrating $\dot M$ and $\dot E$ in Table.\ref{tab:ph_LL_tune} over time. These values are broadly consistent with the general picture of TDEs, in which a stellar-mass object is tidally disrupted by a supermassive black hole and drives a high-velocity outflow with characteristic speeds of $\sim 1.0\times10^{9}\,\mathrm{cm\,s^{-1}}$.  
The estimated kinetic energy is also comparable, in order of magnitude, to the kinetic energy inferred for jet-associated TDEs \citep{cendes2022mildly}.

Figure~\ref{fig:time_evol}(A) shows that the mass-outflow rate in the optically thin outflow decreases approximately following $\propto t^{-2.5}$.
This decline is steeper than the canonical time dependence of the debris fallback rate in TDEs, $\dot{M}_{\rm fb} \propto t^{-5/3}$. 
Assuming that the optically thin component ultimately originates in the fallback material, the result may indicate time-dependent conversion efficiency in either of the fallback-to-accretion or accretion-to-fallback (or even fallback-to-outflow).

The derived values of $\dot M$ are quite high, initially $\approx 0.7 M_\odot$ yr$^{-1}$ that declined to $\sim 0.2 M_\odot$ yr$^{-1}$ in $\sim 100$ days. For comparison, the Eddington accretion rate of the SMBH that triggered ASASSN-14ae is $\approx 0.14~M_{\odot}~{\rm yr}^{-1}$ \citep{caoLimitsLuminosityMass2015}, adopting $M_{\rm BH} \approx 10^{6.8}~M_{\odot}$ \citep{holoien2016six}. 
This comparison suggests that the mass flow associated with the optically thin component is in the regime expected for super-Eddington accretion-driven outflows
\citep[e.g.,][]{yoshiokaRadiationOutflowProperties2024}; note that the super-Eddington accretion means that the accretion disk is fed by the super-Eddington rate from outside, i.e., the fallback rate in this case. In such a case, the excessive material is ejected by a strong outflow before or during the disk accretion. Given that the outflow rate already exceeds the Eddington rate, we suggest this outflow is a possible origin of the optically thin component.  

The kinetic-energy injection rate, shown in Fig.~\ref{fig:time_evol}(B), follows a power-law decline following approximately $\dot E \propto t^{-4.5}$. 
\citet{yoshiokaRadiationOutflowProperties2024} found, based on two-dimensional radiation-hydrodynamic simulations of super-Eddington accretion flows and associated outflows, that the kinetic-energy injection rate is related to the normalized mass accretion rate as $\dot{E} \propto \dot{m}^{2.7}$ for $\dot{m}<400$.
Here, $\dot{m}$ is defined as
\begin{equation}
    \dot{m} \equiv \frac{\dot{M}}{L_{\rm Edd}/c^2}, 
\end{equation}
where $\dot M$ is the rate at which the accretion disk receive the mass from outside, i.e., the fallback rate in the present situation. 
For ASASSN-14ae, the condition $\dot{m}<400$ corresponds to $\dot{M}<5.6~M_{\odot}~{\rm yr}^{-1}$, assuming $M_{\rm BH} \approx 10^{6.8}~M_{\odot}$ as estimated by \citet{holoien2016six}. 

If the disk is fed by the canonical fallback-rate evolution, $\dot{m} \propto t^{-5/3}$, the relation $\dot{E} \propto \dot{m}^{2.7}$ predicts $\dot{E} \propto t^{-4.5}$ -- an excellent match to what we found for ASASSN-14ae. This again reinforces the origin of the optically thin component as an accretion-driven outflow.

Another interesting behavior of the kinetic energy injection rate to the optically thin outflow component is seen in its relation to the evolution of the radiation (bolometric) luminosity. As mentioned in Section~\ref{sec:result_thin}, these two energy budgets roughly overlap, both in the absolute values and in their evolution; given various uncertainties in our model, this similarity may point to a physical connection between the two energy reservoirs.
This suggests that the kinetic energy and the radiation/thermal energy contents are close to equipartition, which motivates the following possible interpretation;
the energy is first channeled into the kinetic energy of the outflow; if this outflow is confined (originally) within the optically thick region, then the equipartition is realized between the kinetic energy and thermal energy contents; once the outflow emerges out of the optically thick region, the thermal energy is quickly radiated away, leading to the comparable amount of the kinetic energy and radiation energy contents. Note that we indeed have the optically thick component in our two component model, which naturally provides a condition for this scenario to operate. 

In this scenario, the underlying assumption is that the dominant `initial' power is in the form of the kinetic energy. This is expected in the super-Eddington accretion case, since the contribution of the radiation in the `initial' channels is limited by the Eddington luminosity. This is in contrast to the case where the accretion rate is much below the Eddington rate, where a strong outflow is not expected and thus the fraction of the energy going into the kinetic energy is far below the radiation energy. 

So, from the analysis of the properties of the optically thin component, we have various indications that this is a strong outflow following the super-Eddington accretion, and this component controls the energy budget of the system. We however note one caveat in this scenario;  ASASSN-14ae appears to be sub-Eddington (at least not extremely super-Eddington) during the optical flare based on the face values of its
bolometric luminosity ($< 10^{44}$ erg s$^{-1}$) and the Eddington luminosity expected from its SMBH mass ($M_{\rm BH} \approx 10^{6.8}~M_{\odot}$ and $L_{\rm Edd} \approx 8.0\times 10^{44}~{\rm erg~s^{-1}}$). This is in tension with the expectation for the super-Eddington accretion, where the radiation luminosity from the accretion disk roughly traces the Eddington luminosity while the excessive energy goes to the kinetic energy of the outflow \citep{yoshiokaRadiationOutflowProperties2024}
However, this may not be necessarily the case when the accretion disk is hidden by the surrounding material (envelope or outflow) and form the photosphere far above the accretion disk, as is inferred for ASASSN-14ae and TDEs in general -- in such a case, the disk-originated radiation will be fully or partially attenuated, and thus the `observed' radiation may not have to trace, and can be lower than, the Eddington luminosity.
We encourage further investigation of the physical processes, including the launch of an outflow and its energetics, in this interesting regime of the accretion rate, including the effect of the envelope and/or outflow that may hide the direct view toward the central system. 

In addition, the interplay between the outflow and optically thick component requires further investigation, including when and how the kinetic energy is converted to thermal energy, and then to radiation energy. We also note the importance of revisiting the SMBH mass estimate -- the value used here was derived based on phenomenological relations like the BH mass -- bulge mass relation \citep{holoien2014asassn14ae}. In other word, our model/scenario may indicate a smaller BH mass than derived previously. Given that one important question to be tackled by studying TDEs is whether various relations derived for AGNs would also hold for non-active SMBH, it is essential to develop a method to independently derive a SMBH mass through TDE observables, and the present work indeed represents such a step. 

It is useful to compare our result with the recent study by \citet{zhangExploringPropertiesPhotosphere2024}, 
who also suggested the need of additional line-emitting materials beside the photospheric component. Our approaches are indeed complementary, and largely independent. The approach by \citet{zhangExploringPropertiesPhotosphere2024} is model/theory-oriented; they modeled properties of photospheres and emission lines based on a slim-disk wind solution and CLOUDY photoionization calculations. They found that a radiation-driven disk wind under-produced H$\alpha$ line luminosity if only the disk-wind material outside the photosphere is considered. They therefore suggested that additional material, possibly supplied by debris self-collision during circularization, may contribute to the line-emitting atmosphere.

At the first look, one might see that their result appears contradicting to the present result, i.e., a single smooth component `overproduces' H$\alpha$. rather than underproduces it. However, their results can indeed be interpreted within our modeling scheme in a consistent manner. The model parameters used in \citet{zhangExploringPropertiesPhotosphere2024} are based on the slim-disk solution that produces $\sim 0.07M_\odot$ of the disk-wind material above the photosphere. This value is about $1/100$ in the `single-component' outflow model in Section \ref{sec:result_epoch_survey}. Therefore, their model, based on the slim-disk model, would result in not only the underproduction of the H$\alpha$ luminosity but also a lack of sufficient materials to keep the extended photosphere. Our model is constructed upon different strategy -- it is more phenomenological, but constrained by observations (e.g., the properties of the photosphere). 
In summary, the differences between our results and \citet{zhangExploringPropertiesPhotosphere2024} stems from the way of the model construction and constraint, and we indeed converge, through different approaches, to the similar conclusion; a need for the multiple components.

\subsection{Origin of the optically thick component}\label{sec:disc_thick}
Our formalism for the optically thick component indeed does not necessarily assume that it is an outflow; since the radial extension of this component is fixed by the observed photospheric radius, the `velocity' of this component is derived by assuming it is an outflow rather than derived directly from observables. 
In fact, several arguments, listed below, are against the interpretation that the optically thick component represents an outflow.

If the optically thick component were described by an outflow, 
materials ejected at earlier epochs should expand toward a larger radius at later times. 
Such outer material should contribute to the optical depth and affect the location of the photosphere. As a result, modeling the optically thick component as an outflow should lead either to the expanding photosphere as a function of time or a large amount of line-forming materials just above the photosphere. 
This behavior is inconsistent with the shrinking photosphere inferred from observations, or with a need of only a moderate amount of materials just above the photosphere for the line formation. This is indeed the main reason of our introducing the two-component model; the separation of the components for the photosphere formation (optically thick component) and the line formation (optically thin component). 

Also, the total mass ejected in the optically thick component is $\gtrsim 2.9 M_\odot$ obtained from Table. \ref{tab:ph_LL_tune}, if it were an outflow. This mass is too much if we assume ASASSN-14ae is an event in which stellar mass object $\sim 1 M_\odot$ is disrupted by SMBH and eject $\leq 0.5 M_\odot$ as an outflow.

To avoid this difficulties, the optically thick component may instead be interpreted in the framework of a quasi-hydrostatic envelope (or a cooling envelope), as proposed by \citet{metzger2022coolinga}. In this scenario,  shocks generated during the circularization of the debris streams dissipate orbital energy and assemble a fraction of the bound debris into a hot, quasi-spherical envelope near the circularization radius. The thermal energy deposited by these shocks provides pressure support, through gas and radiation pressure, against gravity. As the envelope radiates its internal energy on a Kelvin--Helmholtz timescale, it contracts and its photosphere recedes. Although we do not solve the hydrostatic structure or the energy equation of such an envelope in the present phenomenological model, the observed shrinking photosphere of ASASSN-14ae is qualitatively consistent with this picture rather than optically thick outflow.
In this scenario, the envelope itself does not necessarily produce line emission, and thus avoiding the problem of the overproduction of the line luminosity. 
The cooling-envelope scenario also provides observational tests beyond the photospheric emission. It predicts a gradual increase in the effective temperature and a declining optical continuum as the envelope contracts, and may allow delayed thermal X-ray emission to emerge once the inner regions become exposed \citep{metzger2022coolinga}.

As discussed above, the optically thick component may indeed be energetically connected to the optically thin component. In this aspect, this envelope may not be simply `cooling', but heated by the outflow on its way propagating through the envelope. As such, further modeling of the envelope scenario may require an outflow-envelope interaction, and we plan to investigate such a scenario further in a separate paper.

\section{Conclusions}\label{sec:conclusions}
In this study, we investigated emission properties based on the wind-driven model, a framework that has been proposed for TDE outflows. 
In addition to the photospheric properties, we computed the resulting H$\alpha$ line luminosities and profiles by ray-tracing radiative transfer calculations. 

We find that a single component is insufficient to simultaneously reproduce the observed photospheric properties and H$\alpha$ line luminosity of ASASSN-14ae -- if the model is to reproduce the photospheric properties, the same model results in a large amount of materials just above the photosphere (due to the smooth density distribution across the photosphere), resulting in overproduction of the H$\alpha$ luminosity. This argues for a two-component structure, composed of an optically thick inner layer and an optically thin outer layer.

Within this two-component model framework, we identify parameter sets that reproduce the observed properties of ASASSN-14ae. 
From the physical parameters derived for the two components, we suggest the following scenario: 
\begin{enumerate}
\item The optically thick component likely represents a quasi-hydrostatic envelope rather than an outflow. 
\item The optically thin component originates in an accretion-driven outflow, potentially in the super-Eddington regime. 
\item These two components are energetically connected. The outflow that originates from the central engine dissipates its kinetic power through its propagation in the envelope. This process results in the equipartition in its kinetic energy at the emergence as an optically thin component with the thermal energy in the optically thick component. 
\end{enumerate}

The above scenario explains various properties derived for the two components, with the overall mass and energy budget involved in this scenario being consistent with the derived properties.
One potential issue is 
that the radiation (bolometric) luminosity is at the sub-Eddington luminosity expected from the host galaxy mass of ASASS-14ae, against the expectation that the radiation luminosity stays around the Eddington luminosity in the Super-Eddington regime. We indeed suggest that this expectation, the radiation luminosity being equal to the Eddington luminosity, may not hold for ASASSN-14ae (or TDEs in general); our investigation indicates existence of massive/dense envelope and/or outflow that can hide the central system, and the observed radiation does not necessarily originate in the Eddington-limited luminosity from the accretion disk.
This issue requires further investigation, including independent BH mass estimate as well as detailed and quantitative construction of models based on the proposed scenario. 

Our investigation involves various simplifications. As pros, our model here can be readily applied to a sample of well-observed TDEs, and this is the direction we will investigate in a separate paper. As cons, the model requires further update in various physical processes to develop a more quantitative and accurate model, including the treatment of scattering, non-LTE. Updated treatments in ionization and excitation, as well as non-LTE effects, are required especially for ions having high-ionization and excitation potentials, i.e., to model He II; it is deferred to future work.
Also, further quantifying a geometric structure is an important step, in order to better understand the mutual relation of the optically thin and thick components. In any case, the present investigation provides a baseline model to investigate the TDE dynamics through their optical/UV emissions, and sets the scene for future sample study and more detailed investigation. 

\begin{acknowledgments}
The authors acknowledge stimulating discussion with K. Uno, Y. Sakurai. HT acknowledges support from JST SPRING grant No. JPMJSP2110. KM acknowledges support from JSPS KAKENHI grant No. JP24KK0070, JP24H01810, and JP23H04894. This research was supported by a grant from the Hayakawa Satio Fund awarded by the Astronomical Society of Japan
\end{acknowledgments}

\appendix
\twocolumngrid
\section{Effects of an asymmetric outflow}\label{sec:geom}
In Section~\ref{sec:result_epoch_survey}, we concluded that a spherically symmetric single-component model that fits to the observed photospheric radius and temperature results in overproduction of the H$\alpha$ line luminosity. One may suspect that this problem could be remedied by introducing an asymmetric outflow, in which the effective line-emitting volume above the photosphere is reduced. 
In this Appendix, we relax the assumption of spherical symmetry and introduce a conical outflow, where the opening angle is taken as an additional parameter.

\subsection{Model Setting}
In considering a conical outflow, we extract a bipolar conical region with half-opening angle $\theta$ from the spherically-symmetric counterpart used in Section~\ref{sec:survey}. We perform radiative transfer calculations within this restricted emitting volume.
We then investigate whether there exists a set of parameters $(\theta, \dot{M}, v)$ that simultaneously reproduces the observed line properties and the photospheric properties.

In the radiative transfer calculations, we adopt the same photospheric criterion as in the spherically symmetric case (see Section~\ref{sec:rad_trans}).
Accordingly, rays originating from regions deeper than the photosphere along the lateral surface of the cone are initialized with Planck emission at the photospheric temperature.

\subsection{Effects of reducing the opening angle}
Starting the same single-component model presented in Section~\ref{sec:survey}, the line luminosity at each epoch is reduced by the solid angle of the outflow as compared to the spherical case. This defines the opening-angle of the conical outflow to bring the predicted line luminosity to the observed level; it is 19, 28.5, 33, 41, 48 and 51 degree for epochs 4, 30, 39, 51, 73 and 94 days. 

The above values for the solid angle are derived for the viewing angle perpendicular to the bipolar axis (i.e., an equatorial view). However, we note that the line luminosity is insensitive to the viewing direction as H$\alpha$ is emission dominated. It is the line profile that is sensitive to the viewing direction, analogous to the late-time nebular-line formation in supernovae \citep[e.g.,][]{maedaOpticalEmissionAspherical2006}. 

\begin{figure}[t]
    \centering
    \includegraphics[width=\columnwidth]{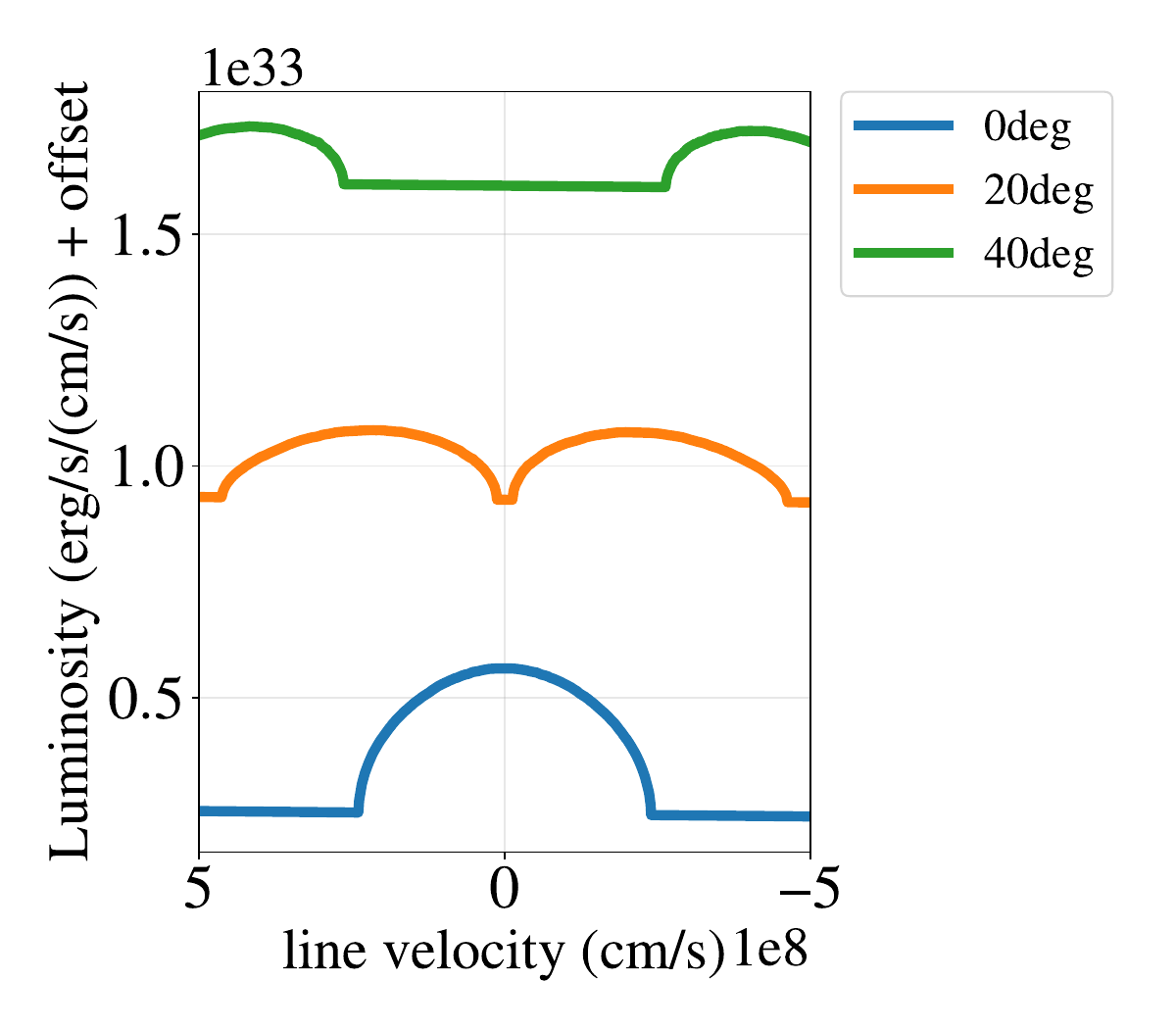}
    \caption{
    Variation of the H$\alpha$ line profile at epoch 4d depending on the viewing direction. 
    The blue, orange, and green curves correspond to viewing angles of 0° (equatorial), 20°, and 40° from the equatorial direction, respectively.
    The profiles are vertically offset for clarity.
    }
    \label{fig:epoch4full}
    
\end{figure}

If the viewing angle is varied, the line profile changes as illustrated in Figure~\ref{fig:epoch4full}.  
When the viewing angle deviates significantly from the equatorial direction, the line profile develops a double-peaked (split) structure that is generally not observed in TDEs.
Therefore, in order to obtain a line profile centered around the rest wavelength, as typically observed, the viewing angle cannot be substantially inclined toward the polar axis. This already provides one argument against a strongly collimated outflow -- the viewing angle is expected to be random, and then we should observe such a double-peaked profile in a fraction of TDEs. 

In addition, for an equatorial viewing angle, the resulting profile exhibits a (much) smaller FWHM than the spherical counterpart with the same outflow velocity, since the line-of-sight velocity is significantly reduced in this case. This leads to the predicted line FWHM much smaller than observed; 
at epoch 4d, the observed $v_{\rm FWHM}$ (the velocity corresponding to the FWHM) is approximately $10^9\,{\rm cm\,s^{-1}}$, while the one predicted by the conical outflow reaches only $\sim 4 \times 10^{8}$ cm s$^{-1}$ (see Figure~\ref{fig:epoch4full} and the first row of Table~\ref{tab:open_v_tune}). 

Then, the FWHM of the model should be broadened by 
increasing the wind velocity $v$. 
This exercise is summarized in Table~\ref{tab:open_v_tune}, where 
the second row corresponds to the parameters adjusted to match the observed $v_{\rm FWHM}$ by increasing $v$. While the density scale ($\dot M / v$) is not changed (as fixed by the photospheric properties), the outflow is now extended to much larger radius (since the outer edge is determined by $R_{\rm out} = {\rm day} \times v$). This increased emitting volume now results in a line luminosity larger than observed. 
To reduce the line luminosity to the observed level ($1.1 \times 10^{41}$ erg s$^{-1}$), we then need to decrease the opening angle further (as shown by the third row) - this then again results in the reduction of the FWHM for an equatorial observer. The exercise above demonstrates the problem -- we are always encountered by a problem either in the line luminosity or FWHM, even introducing a conical outflow.

\begin{deluxetable*}{lccccc}
\tablecaption{Results of tuning of line luminosity and FWHM at epoch 4d in ASASSN-14ae \label{tab:open_v_tune}}
\tablehead{
\colhead{$v$ ($10^8$cm/s)} &
\colhead{$\dot{M}$ ($M_\odot$ yr$^{-1}$)} &
\colhead{$v_{\mathrm{FWHM}}$ ($10^8$cm/s)} &
\colhead{open angle (degree)} &
\colhead{H$\alpha$ Line Luminosity ($10^{41}$erg/s)} 
}
\startdata
7.3 & 9.94 & 4.1 & 19 & 1.2 \\
20.0 & 25.0 & 11.0 & 19 & 6.9\\
20.0 & 25.0 & 4.8 & 6.1 & 1.2 \\
\enddata
\end{deluxetable*}

In summary, introducing an asymmetric outflow would not solve the problem. 
Although the line luminosity can be adjusted in the single-component model by introducing a finite opening angle, this modification predicts that a fraction of TDEs should show double-peaked line profiles\footnote{A unipolar model would also result in the shift in the wavelength, and thus would not solve the problem.}, which are not typically observed in TDEs.
Moreover, attempting to match the observed FWHM leads to a dilemma; increasing the wind velocity broadens the profile but results in overproduction of the line luminosity; compensating for the excess luminosity then requires a smaller opening angle, which in turn reduces the FWHM again. This feedback prevents the simultaneous reproduction of all the observational properties.

\section{Single component parameter survey in other epochs}\label{sec:other_epochs_single}
\begin{figure}[htbp]
    \centering
    \includegraphics[width=\columnwidth]{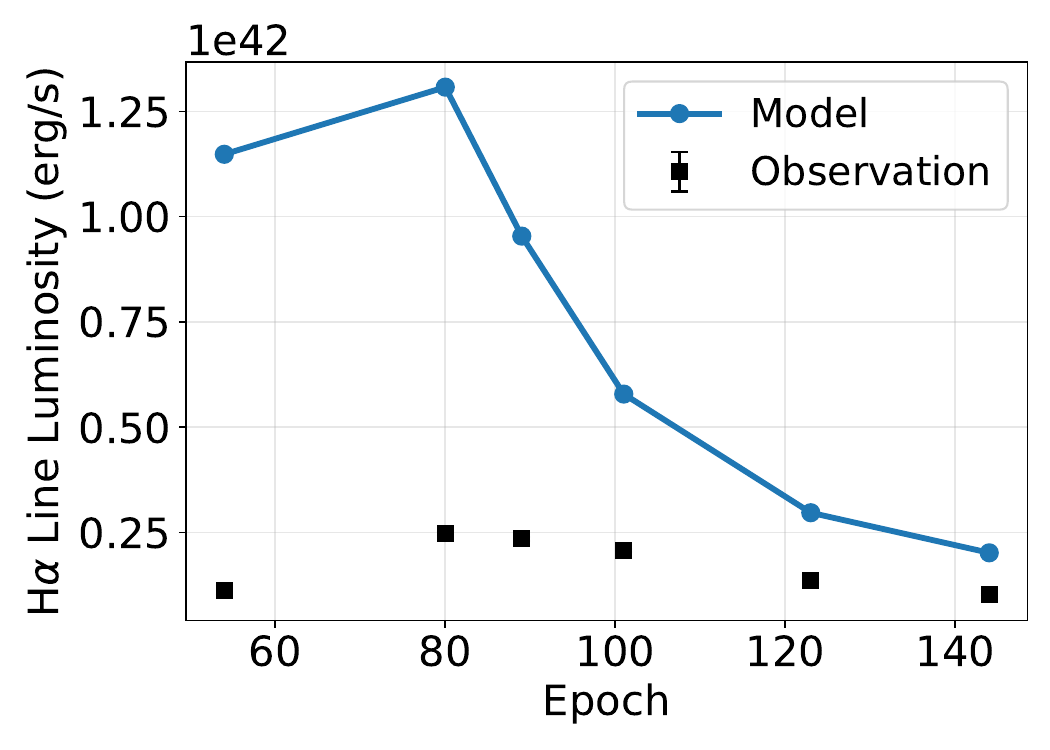}
    \caption{
    Results of the $\dot{M}$ survey at all epochs in ASASSN-14ae, based on the one-component model.
    Black points show the observed H$\alpha$ line luminosity at each epoch, and blue points show the calculated H$\alpha$ line luminosity by the model constrained by the observed photospheric radius ($R\ph$) and temperature ($T\ph$). 
    }
    \label{fig:epoch_every}
\end{figure}

Figure~\ref{fig:epoch_every} shows the results of the $\dot M$ survey for the one-component model conducted for all epochs. It shows that the overproduction problem of the H$\alpha$ luminosity exists at all epochs -- once the (one-component) model is fixed by the photospheric properties, the model does not reproduce the H$\alpha$ luminosity.  

\section{Line profile of the Two-component model in all epochs}\label{sec:other_TC}
\begin{figure}[htbp]
\centering
\begin{subfigure}{0.45\columnwidth}
    \centering
    \includegraphics[width=\linewidth]{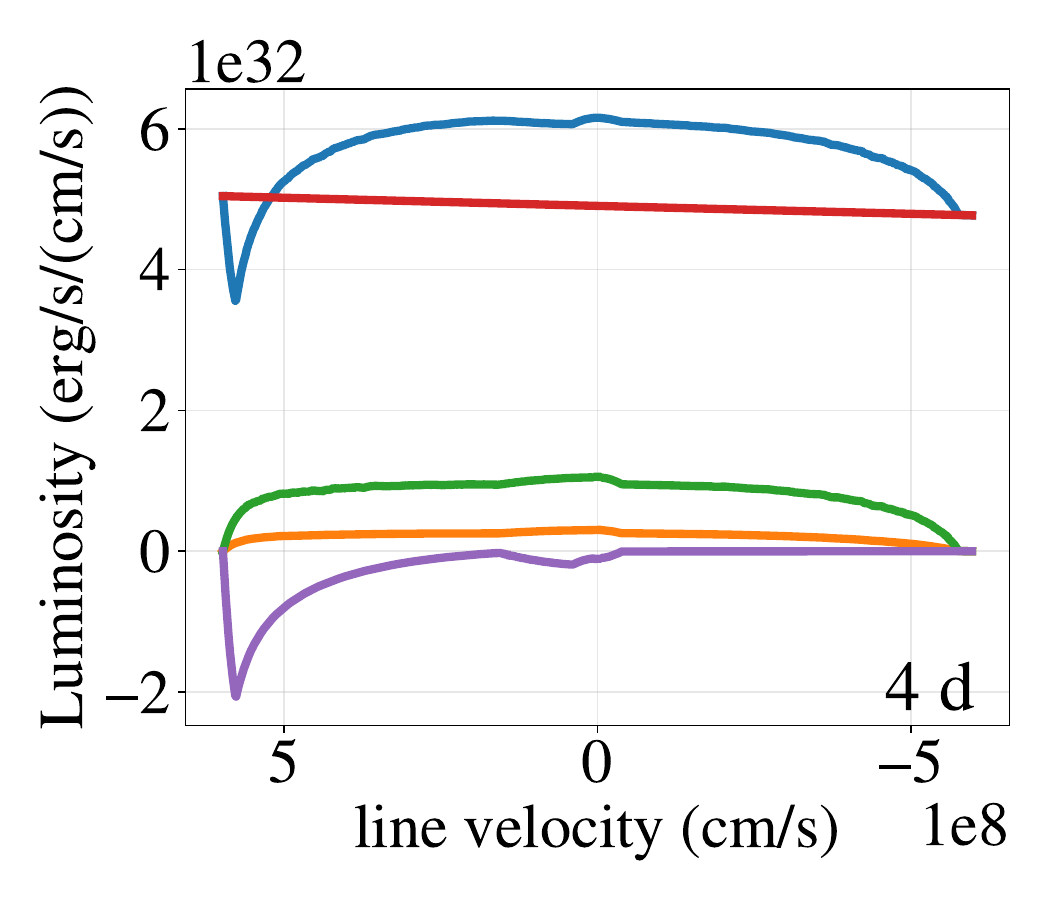}
\end{subfigure}
\hfill
\begin{subfigure}{0.45\columnwidth}
    \centering
    \includegraphics[width=\linewidth]{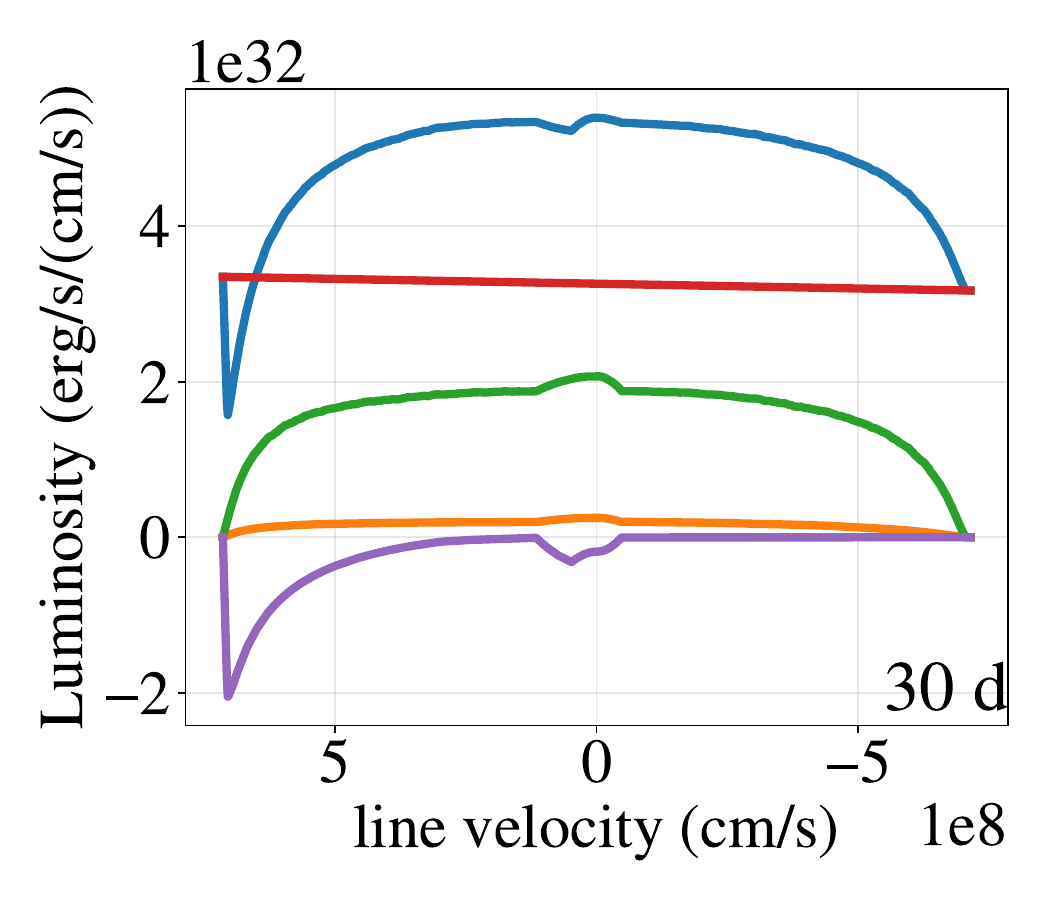}
\end{subfigure}

\vspace{2mm}

\begin{subfigure}{0.45\columnwidth}
    \centering
    \includegraphics[width=\linewidth]{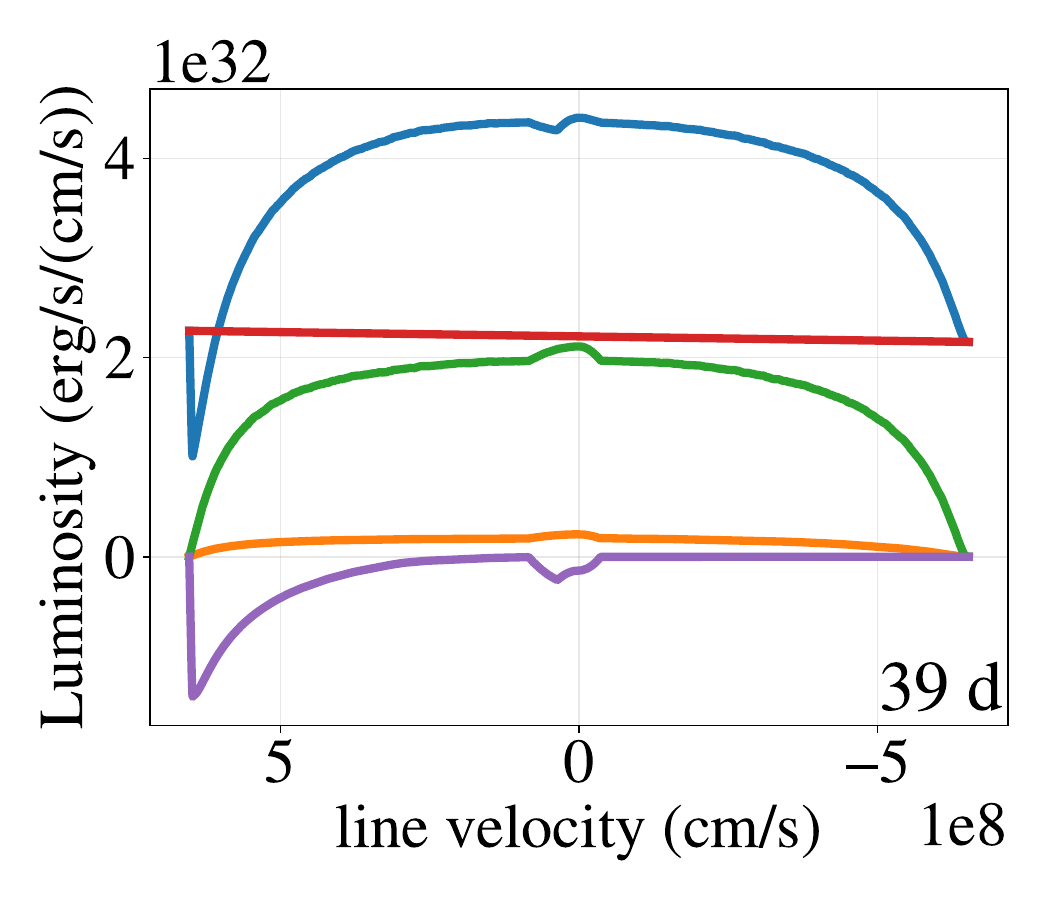}
\end{subfigure}
\hfill
\begin{subfigure}{0.45\columnwidth}
    \centering
    \includegraphics[width=\linewidth]{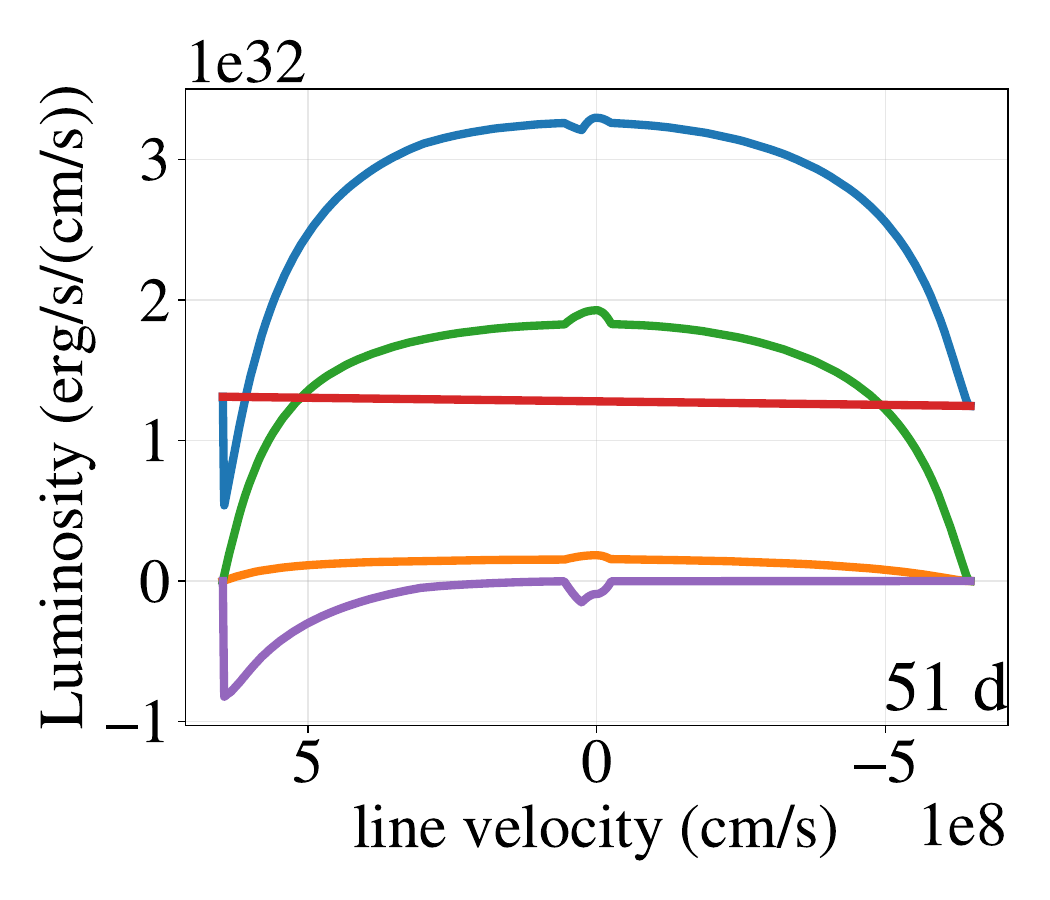}
\end{subfigure}

\vspace{2mm}

\begin{subfigure}{0.45\columnwidth}
    \centering
    \includegraphics[width=\linewidth]{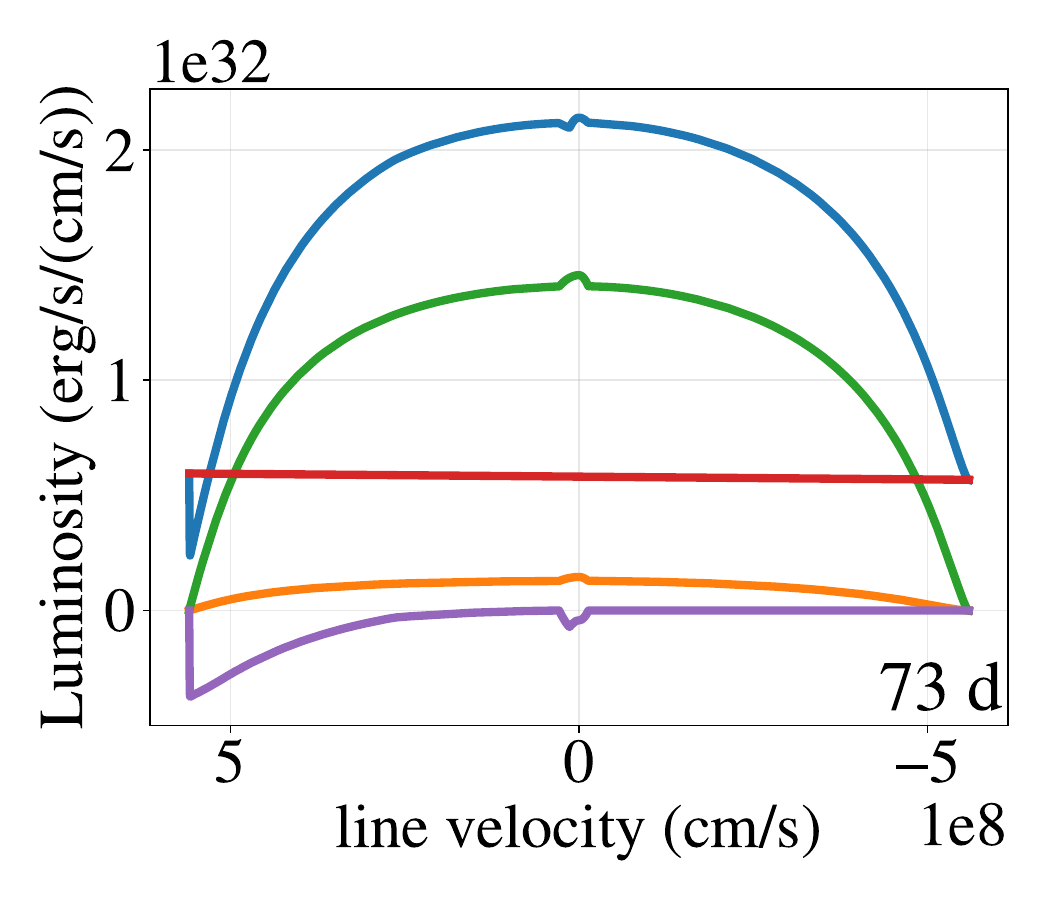}
\end{subfigure}
\hfill
\begin{subfigure}{0.45\columnwidth}
    \centering
    \includegraphics[width=\linewidth]{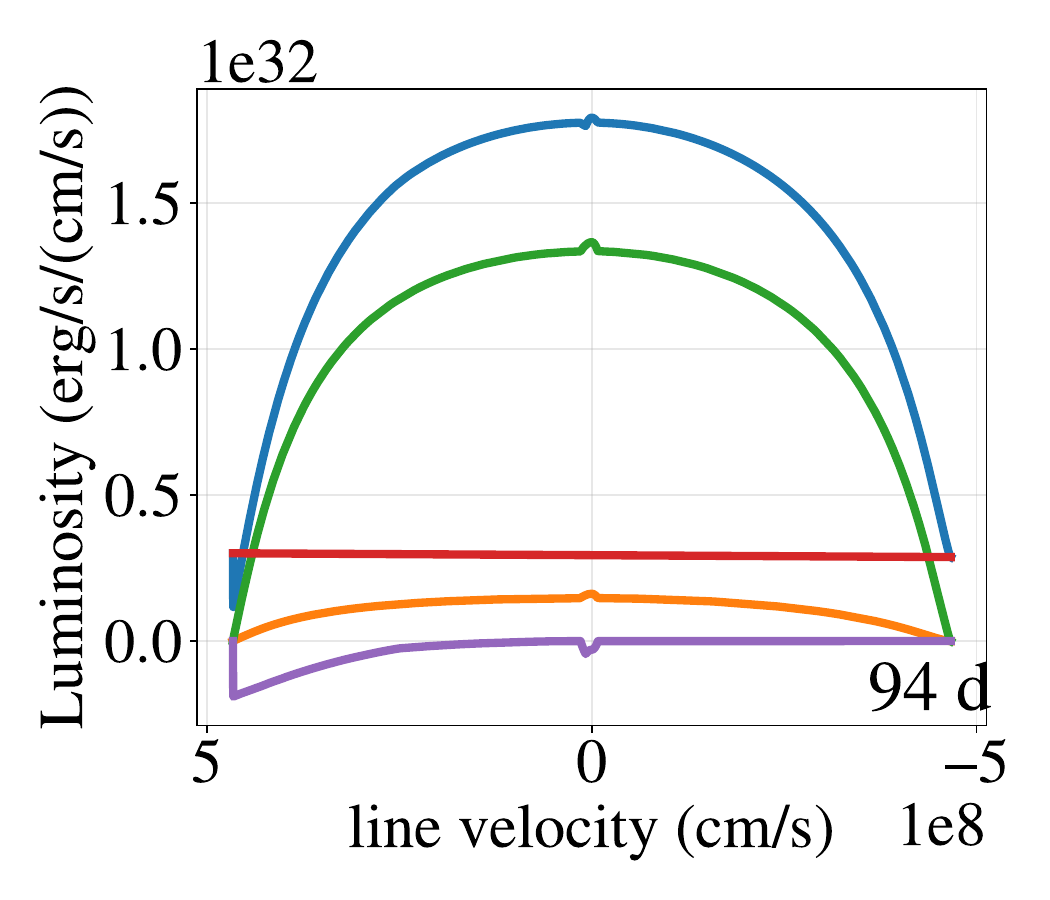}
\end{subfigure}

\caption{Time evolution of the H$\alpha$ line profiles in the two component model. 
}
\label{fig:lp_double_outflow_2x3_singlecol}

\end{figure}

Figure~\ref{fig:lp_double_outflow_2x3_singlecol} shows the results of the line profile calculations for all epochs (see Section~\ref{sec:LP_TC}).

\section{Validity and limitations of the LTE/Saha treatment}
\label{sec:lte_validity}
In our calculations, we obtain the hydrogen ionization fraction from the Saha equation and the level populations from the Boltzmann distribution. This LTE treatment is an approximate closure for the ionization and excitation states, not a substitute for a non-LTE rate-equation calculation. In particular, the radiation field in the optically thin line-forming region is geometrically diluted and need not be Planckian at the local gas temperature. We therefore address two distinct questions: whether the ionization state can respond faster than the outflow expands, and how much possible departure from LTE would affect the inferred H$\alpha$ emissivity. 

The local expansion timescale is
\begin{equation}
    t_{\rm dyn} = \frac{r}{v} .
\end{equation}
We estimate the Case B recombination timescale as
\begin{equation}
    t_{\rm rec} = \frac{1}{\alpha_{\rm B}(T)n_e} ,
\end{equation}
where $\alpha_{\rm B}(T)$ is the Case B recombination coefficient evaluated at the local gas temperature. The photoionization timescale is
\begin{equation}
    t_{\rm ph} = \Gamma_{\rm ph}^{-1} ,
\end{equation}
where the photoionization rate per neutral hydrogen atom is
\begin{equation}
    \Gamma_{\rm ph}(r)
    = 4\pi \int_{\nu_0}^{\infty}
    \frac{J_\nu(r)}{h\nu}\,\sigma_\nu\,{\rm d}\nu .
\end{equation}
Here, $\nu_0$ is the hydrogen ionization-threshold frequency and $\sigma_\nu$ is the ground-state bound-free cross section. We approximate the radiation field as a geometrically diluted photospheric blackbody,
\begin{equation}
    J_\nu(r) = W(r) B_\nu(T_{\rm ph}) ,
\end{equation}
where
\begin{equation}
    W(r) = \frac{1}{2}
    \left[1-\sqrt{1-\left(\frac{R_{\rm ph}}{r}\right)^2}\right] .
\end{equation}
Adopting the hydrogenic approximation $\sigma_\nu=\sigma_0(\nu_0/\nu)^3$, this expression becomes
\begin{equation}
\begin{split}
    \Gamma_{\rm ph}(r)
    ={}& 4\pi
    \left[1-\sqrt{1-\left(\frac{R_{\rm ph}}{r}\right)^2}\right]
    \frac{\sigma_0\nu_0^3}{c^2} \\
    &\times \int_{x_0}^{\infty}\frac{{\rm d}x}{x(e^x-1)} ,
\end{split}
\end{equation}
where $x_0=h\nu_0/(k_{\rm B}T_{\rm ph})$. This photoionization rate is not used to set the ionization fraction in our fiducial model; it is used here to only measure how rapidly the gas would respond to the adopted photospheric radiation field. Moreover, both $n_e$ and the radiation field entering this estimate are taken from the fiducial LTE/Saha solution. The comparison is therefore an a posteriori consistency check, not an independent non-LTE solution.

\begin{figure}[t]
    \centering
    \includegraphics[width=\columnwidth]{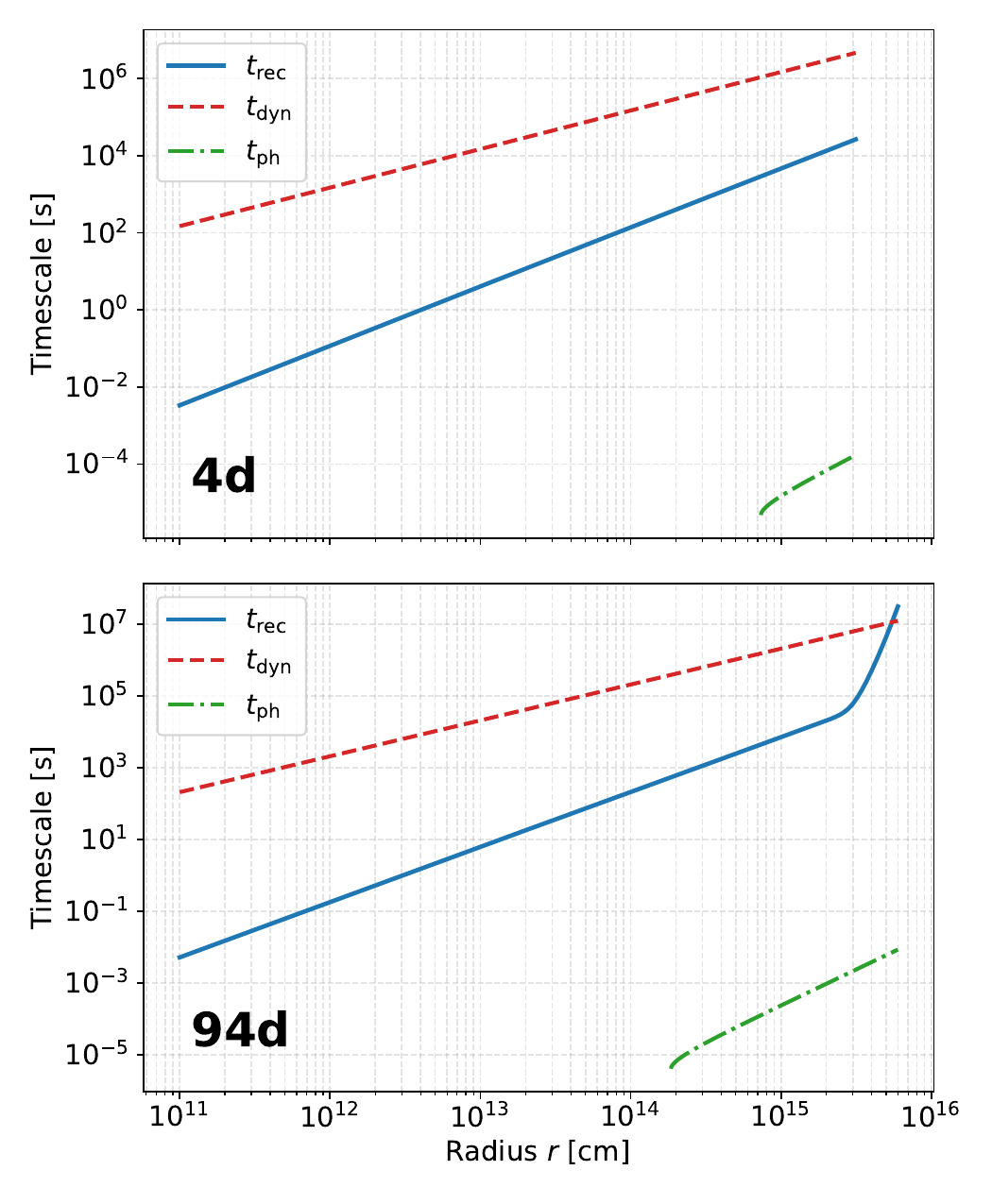}
    \caption{
    A posteriori comparison of the recombination, photoionization, and expansion timescales.
    The electron density and radiation field used here are taken from the fiducial LTE/Saha model.}
    
    \label{fig:ion_timescales}
\end{figure}

Figure~\ref{fig:ion_timescales} shows that $t_{\rm ph}<t_{\rm dyn}$ in the entire spatial domain at both epochs. At 4 days, $t_{\rm rec}<t_{\rm dyn}$ everywhere, whereas at 94 days $t_{\rm rec}$ approaches and exceeds $t_{\rm dyn}$ only near the outer boundary. These outermost layers contribute little to the integrated H$\alpha$ luminosity because of their low density. The exercise therefore shows that the strong coupling between the gas and radiation should be realized under the situation considered in the present work, fulfilling one important condition for LTE. 
Furthermore, the bulk of the modeled line emission is not expected to suffer strong time-dependent ionization freeze-out, fully justifying the use of an instantaneous ionization closure. 

While the timescale argument above suggests that LTE is a good approximation, it does not fully reject departure from LTE. 
Strict LTE is unlikely to hold throughout an optically thin, expanding atmosphere, and non-LTE effects can alter both the neutral fraction and the excited-state populations. Detailed calculations of expanding supernova atmospheres likewise demonstrate that a non-LTE treatment is sometimes required to determine ionization states and line formation reliably \citep[e.g.,][]{baronNonlocalThermodynamicEquilibrium1996}, although their physical conditions differ from those in a TDE outflow and do not provide a unique correction factor for the present model. To quantify the sensitivity of our inference, we write the actual H$\alpha$ emissivity as
\begin{equation}
    j_{\rm H\alpha} = f_{\rm NLTE}\,j_{\rm H\alpha}^{\rm LTE}
\end{equation}
and adopt $1/3\leq f_{\rm NLTE}\leq 3$ as a conservative systematic bracket. This range is a sensitivity test rather than the result of a self-consistent non-LTE calculation.

In the optically thin, resonance-scattering-dominated limit, $L_{\rm H\alpha}\propto \rho\propto\dot{M}$. The mass-outflow rate inferred from a fixed observed H$\alpha$ luminosity therefore scales approximately as $\dot{M}\propto f_{\rm NLTE}$. Our adopted bracket corresponds to a systematic uncertainty of a factor of $0.33$--$3.0$, or approximately $\pm0.48$ dex, in the H$\alpha$-based density normalization. 
For the single component outflow modeling (in Section~\ref{sec:result_epoch_survey}), the mass-outflow-rate $\dot M$ derived through the photospheric constraints gives larger H$\alpha$ emission than observation, by more than a factor of $\sim 5$ in the earlier phases (see also Appendix~\ref{fig:epoch_every}). 
The discrepancy is larger toward the earlier phase, while non-LTE effects are expected to become stronger toward the later, low-density regime. 
Non-LTE effects may shift the quantitative parameters at the factor-of-a-few level, but they are unlikely to remove the need for a substantial density contrast between the photospheric and line-forming regions. A self-consistent non-LTE calculation will nevertheless be required for more precise parameter estimates and for simultaneous modeling of multiple hydrogen and helium lines.

\FloatBarrier

\end{document}